\documentclass[monophys,vecphys,envcountchap,natbib]{svmono}

\usepackage{makeidx}         % allows index generation
\usepackage{graphicx}        % standard LaTeX graphics tool
\usepackage{multicol}        % used for the two-column index
\usepackage[bottom]{footmisc}% places footnotes at page bottom

\usepackage{amsmath}
\usepackage{amssymb}

\usepackage{bbm}
\usepackage{nicefrac}
\usepackage{mathrsfs}

\usepackage{exscale}      % Skaliert Mathe-Modus Ausgaben in allen Umgebungen richtig.
\usepackage{theorem}      % ermöglicht zusaetzliche Konfiguration von Theorem-Umgebungen, u.a. \theoremstyle
\usepackage{feynmf}
\usepackage{txfonts}
\newfont{\suet}{suet14}
\newfont{\schwell}{schwell}

\begin{document}
\setlength{\unitlength}{1mm}

\thispagestyle{empty}
\author{P. Strack}
\title{Kinetics of Oscillating Neutrinos}

\subtitle{\vspace{45mm}
\begin{center}
\Large{A Master's Thesis Submitted to the Faculty of the\\
DEPARTMENT OF PHYSICS\\
In Partial Fulfillment of the Requirements For the Degree of\\
MASTER OF SCIENCE\\
In the Graduate College\\
The University of Arizona\\
(Accepted and defended April 2005)\\[15mm]
Copyright  \copyright\hspace{2mm}Philipp Strack 2005}
\end{center}}

\maketitle

\frontmatter%%%%%%%%%%%%%%%%%%%%%%%%%%%%%%%%%%%%%%%%%%%%%%%%%%%%%%

\preface

This Master's thesis has been submitted in partial fulfillment of
requirements for an advanced degree at The University of Arizona and
is deposited in the University Library to be made available to
borrowers under the rules of the Library.\vspace{3mm}

Brief quotations from this Master's thesis are allowable without
special permission, provided that an accurate acknowledgment of
source is made. Requests for permission for extended quotation from
or reproduction of this manuscript in whole or in part may be
granted by the copyright holder.\vspace{3mm}

SIGNED:\hspace{3mm} \textit{Philipp Strack}

\acknowledgements

I thank the selection committee of the \textit{Akademische
Auslandsamt der Universit\"at Heidelberg} for providing me with this
wonderful opportunity to study at the Departments of Astronomy and
Physics of the University of Arizona. The Heidelberg professors Dr.
J. Stachel and Dr. C. Wetterich have helped me with the application
for my studies in the USA and I thank them. I acknowledge financial
and organizational support from the German Fulbright-Commission, and
the Baden-Wuerttemberg Stiftung. They have made life
easier.\vspace{3mm}

I am indebted to my supervisor Prof. Dr. Adam Burrows for
intense\footnote{Efforts leading to this thesis were started at the
beginning of September 2004 and finalized by the end of March 2005
and were at all times accompanied by three graduate courses in
physics.} and superb scientific guidance. This collaboration has
been an honor for me.\vspace{3mm}

The main part of this thesis, the generalized Boltzmann formalism in
flat spacetime is published in Physical Review D (Strack \& Burrows
2005). We acknowledge support for this work through the SciDAC
program of the Department of Energy under grant number
DE-FC02-01ER41184. In particular, we thank Prof. Dr. Z. Chacko for
reading the manuscript of the paper and for useful discussions, and
C. Meakin and M. Prescher for helpful conversations and for reading
the manuscript of this thesis. The latter has also provided
invaluable help with technicalities and latex and has thus softened
the author's inability to use computers.\vspace{3mm}

The members of my thesis committee consisting of the professors Dr.
A. Burrows, Dr. Z. Chacko and Dr. J. P. Rutherfoord, gave generously
of their time to help me for which I thank them wholeheartedly. I
express my gratitude to Prof. Dr. W. J. Duschl who has supported me
more than once with applications.\vspace{3mm}

This work was carried out at Steward Observatory and I gratefully
acknowledge the kind hospitality I have received. The graduate
students here at Steward Observatory have been very integrative and
as representatives for all of them I wish to thank R. S. Bussmann,
I. Momcheva, H. J. Seo, N. Siegler, S. Sivanandam, A.
Diamond-Stanic, and J. Trump.\vspace{3mm}

My whole family and in particular my father, Rolf Dieter Strack, and
my beloved partner in life, Anne Gerrit Knepel, have showered me
with love. I hope I can return this one day. Gerrit has also
proof-read the whole thesis and I thank her.
\newpage
\thispagestyle{empty} \vspace{3.5cm}
\begin{flushright}
% write your text here
{\Large Dedicated to my grandfather Georg Walter Strack (1923 --)}
\end{flushright}
\abs \vspace{-6mm} In the standard approach to neutrino transport in
the simulation of core-collapse supernovae, one calculates the
spatial, temporal, and spectral evolution for each neutrino species
in terms of the classical density in phase-space or the associated
specific intensity. The neutrino radiation is coupled to matter by
source and sink terms on the ``right-hand-side" of the transport
equation and together with the equations of hydrodynamics this set
of coupled partial differential equations describes, in principle,
the evolution of core collapse and explosion. However, with the
possibility of neutrino oscillations between species, a purely
quantum-physical effect, how to generalize this set of classical
Boltzmann equations to reflect oscillation physics has not been
clear. To date, the formalisms developed have retained the character
of quantum operator physics and have not been suitable for easy
incorporation into standard supernova codes. In this thesis, we
derive generalized Boltzmann equations for quasi-classical,
real-valued phase-space densities with which any numerical code for
the classical transport equation can easily be generalized to
include neutrino oscillations. This new, extended set of equations
incorporates the full mixing Hamiltonian consisting of the vacuum
oscillations, the matter-induced effective mass for the $\nu_{e}$,
and the self-interactions from neutrino-neutrino scattering. The
equations are intrinsically nonlinear for two reasons: blocking
correction at high densities and the off-diagonal $\nu$-$\nu$
Hamiltonian which maintains SU(N) flavor symmetry for an N-flavor
system. Standard oscillation phenomenology including resonant flavor
conversion (the MSW effect) and the interplay between decohering
matter coupling and flavor oscillations are retained in a
quantum-physically consistent fashion. The extension of the
formalism to curved spacetime is discussed. One has to take into
account two modifications. Firstly, spinors in curved spacetime must
be analyzed with the spin connection in a noncoordinate basis and,
secondly, for the kinetic terms on the left-hand side of the
Boltzmann equation the Christoffel connection must be employed. Both
changes are brought to an algebraic footing via vielbein (tetrad)
fields associated with the square root of the metric. The
generalized kinetic equations in curved space derived in this work
describe neutrino evolution in the most general setting encountered
in astrophysical contexts such as the early universe or in the
vicinity of supernova cores.
\tableofcontents

%%%%%%%%%%%%%%%%%%%%%%%%%%%%%%%%%%%%%%%%%%%%%%%%%%%%%%%
\mainmatter

\chapter[Introduction]{Introduction}
Mixed particles and their oscillations are fundamental for a wide
range of interesting physics: quark-mixing by the
Cabbibo-Kobayashi-Maskawa matrix, its leptonic analog for massive
neutrinos (Bilenky \& Petcov 1977 and 1987), hypothetical
photon-axion and photon-graviton oscillations in the presence of
external magnetic fields (Moertsell, Bergstroem \& Goobar 2002,
Raffelt \& Stodolsky 1988) and $K^{0}-\bar{K}^{0}$ oscillations
(Akhmedov, Barroso \& Keraenen 2001).\\

Physically, these quantum systems are coupled to macroscopic systems
and through external interaction their quantum evolution is altered.
One prominent astrophysical context in which such oscillation for
macroscopic systems is important, involves the neutrinos in
supernova cores that may execute flavor oscillations while
simultaneously interacting with ambient supernova matter (Takahashi,
Sato, Burrows \& Thompson 2003, Thompson, Burrows \& Horvath 2000,
Burrows \& Gandhi 1995, Raffelt 1996, Pantaleone 1995, Qian \&
Fuller 1995, and Wolfenstein 1979).\\

The primary motivation of this work is to provide a straightforward
generalization of the Boltzmann formalism with which to analyze the
kinetics of oscillating neutrinos with collisions including energy
redistribution and inelastic scattering. By taking ensemble-averaged
matrix elements of quantum field operators for mixed particles,
following the pioneering work in the references (Raffelt \& Sigl
1993, Raffelt, Sigl \& Stodolsky 1992, Akhiezer \& Peletminskii
1981, and de Groot, van Leeuwen \& van Weert 1980), we obtain
quasi-classical phase-space densities that satisfy simple Boltzmann
equations with coupling terms that account for the neutrino
oscillations. The formalism is clear, numerically tractable, and
does not contain operators or wavefunctions.\\

Finally, we discuss how to extend our formalism to curved spacetime.
This includes the treatment of generalized resonances similar to the
MSW effect. We present kinetic equations for oscillating neutrinos
interacting with a background medium in curved spacetime.

The plan of this thesis is as follows.\\

\begin{itemize}

\item In chapter \ref{wigner}, we introduce the Wigner phase-space
density operator approach from which we derive our formalism
involving classical phase-space neutrino flavor densities and their
off-diagonal, overlap correlates. The latter couple the different
flavor states to account for neutrino oscillations.\\

    \item In chapter \ref{quasi}, the different
contributions to the mixing Hamiltonian originating in the vacuum
oscillations, matter-induced effective mass for the $\nu_{e}$, and
the neutrino-neutrino scattering are discussed. The full set of
Boltzmann equations is derived for both the flavor phase-space
densities and the associated specific intensities. The advantage of
the specific intensity representation is that collision terms can be
written in a simple and intuitive fashion.\\
    \item In chapter \ref{decoherence}, we demonstrate the usefulness and applicability
of our formalism by successfully testing it on its
quantum-mechanical accuracy. We demonstrate with several simple
examples that the set of equations reproduces what one would expect
from the standard quantum mechanics wavefunction approach.

Firstly, we give analytic solutions to the standard vacuum flavor
oscillations for neutrinos in a box. Secondly, we consider flavor
oscillations with absorptive matter coupling. This is frequently
referred to as quantum decoherence (Raffelt, Sigl \& Stodolsky
1993). Thirdly, it is shown that our formalism is capable of
calculating the neutrino evolution in phase-space when the resonant
matter-induced flavor conversion (MSW effect) is important. We show
complete analytic and numerical congruence to the existing
wavefunction formalism as discussed in the papers by Wolfenstein
1978, Bethe 1986, Mikheyev \& Smirnov 1986 .

We show that our formalism can provide a tool to analyze
active-sterile neutrino conversion in a beam with absorptive matter
coupling. One can observe ``indirect decoherence'' of the
active-sterile conversion. This means that even though the sterile
neutrinos do not interact with a medium their oscillatory pattern is
indirectly influenced since the $\nu_{e}$'s execute
absorptive collisions.\\

    \item In part \ref{curved}, we sketch how to make the relevant changes
    if one wishes to apply our formalism in a curved spacetime. This is potentially important
    in astrophysical contexts such as the early universe and in the vicinity of supernova
    cores.\\

    \item In chapter \ref{pheno}, we
    analyze in a heuristic fashion how neutrino oscillation phenomenology is modified in curved space.
    Therefore, we introduce the vielbein formalism which is the necessary ingredient to
    do calculations involving spinors in the framework of the general theory of
    relativity. Furthermore the vielbein fields are useful tools to write the Liouville
    operator on the left-hand-side of the Boltzmann equation in a
    manifestly covariant fashion to account for the curvature.
    This is done in section \ref{curvedboltz}.\\

    \item Culminating in our efforts, we present new kinetic equations for oscillating neutrinos
    with collisions in curved
    spacetime in chapter \ref{curvedformalism}. The equations of section \ref{sec:gravkin} include the
    possibility of spin-flips into a
    sterile right-handed species by
    coupling of the anomalous magnetic moment of the neutrinos to an
    external magnetic field. In principle, eqs. (\ref{eq:curvedboltz}) govern the evolution of the
    quasi-classical density in phase-space for interacting, massive
    neutrinos including all known effects -- from the quantum-mechanical
    viewpoint as well as from the gravitational perspective.\\

\end{itemize}

\part{Flat Spacetime}

\chapter[Wigner Phase-Space Density Operator]{Wigner Phase-Space Density Operator}
\label{wigner}
\section{Construction}
To analyze a multi-particle system and reflect its inherent
statistics, one usually sets up the density matrix of the system
(Akhiezer \& Peletminskii 1981, de Groot, van Leeuwen \& van Weert
1980). For classical systems this is the phase-space density. For
quantum fields, the conceptually most similar analog is the Wigner
phase-space density operator (Wigner 1932 \& 1984). In what follows
we shall briefly sketch how to construct this quantity.\\

For Dirac neutrinos, we begin with the Dirac equation
\begin{eqnarray}
i(\gamma^{\mu}\partial_{\mu}-m)\psi=0\,\,,
\end{eqnarray}
which describes the dynamics of spin-$\frac{1}{2}$ fields, $\psi$,
in Minkowski space-time with the metric
\begin{eqnarray}
\eta^{\mu\nu}= \text{diag}\,(1,-1,-1,-1)\,\,.
\end{eqnarray}
$\gamma^{\mu}$ are Dirac matrices, which satisfy
\begin{eqnarray}
\{\gamma^{\mu},\gamma^{\nu }\}= 2\eta^{\mu\nu}. \label{diracalgebra}
\end{eqnarray}
Curly brackets denote matrix anti-commutators. Following the
standard second quantization procedure, we write for the momentum
eigenstate expansion of general Dirac fermion annihilation operator
in position space:
\begin{eqnarray}
\psi(\mathbf{r},t)&=&\int \frac{d^{3}p}{(2\pi)^{3}}
\sum_{s}a^{s}_{\mathbf{p}}u^{s}(\mathbf{p},E_{\mathbf{p}})
e^{-i\mathbf{p}\mathbf{r}} e^{-iE_{\mathbf{p}}t}+b^{s
\dag}_{\mathbf{p}}
v^{s}(\mathbf{p},E_{\mathbf{p}})e^{i\mathbf{p}\mathbf{r}}
e^{iE_{\mathbf{p}}t}\,\,,
\end{eqnarray}
and for the Dirac fermion creation operator in position space:
\begin{eqnarray}
\psi^{\dag}(\mathbf{r},t)=\int \frac{d^{3}p}{(2\pi)^{3}}
\sum_{s}b^{s}_{\mathbf{p}}v^{s}(\mathbf{p},E_{\mathbf{p}})
e^{-i\mathbf{p}\mathbf{r}} e^{-iE_{\mathbf{p}}t}+a^{s
\dag}_{\mathbf{p}}
u^{s}(\mathbf{p},E_{\mathbf{p}})e^{i\mathbf{p}\mathbf{r}}
e^{iE_{\mathbf{p}}t}\,\,,
\end{eqnarray}
where $u^{s}(\mathbf{p},t)$ and $v^{s}(\mathbf{p},t)$ are
four-component Dirac spinors describing fermionic and antifermionic
states, respectively. The summation is over additional quantum
numbers, e.g. spin states of the particle. We now tailor
$\psi(\mathbf{r},t)$ to the physical properties of the neutrino and
define
\begin{eqnarray}
\psi (\mathbf{r},t)=\int \frac{d^{3}p}{(2\pi)^{3}}
\left[a_{\mathbf{p}}(t) u(\mathbf{p},E_{\mathbf{p}})+
b^{\dag}_{\mathbf{-p}}(t)
v(\mathbf{p},E_{\mathbf{p}})\right]e^{i\mathbf{p}\mathbf{r}}\,\,.
\label{nuop}
\end{eqnarray}
\\

A Fourier transformation connects momentum and position space
representation of the operators:
\begin{eqnarray}
a(\mathbf{p},t)&=&\int \frac{d^{3}\mathbf{r}}{(2\pi)^{3}}u^{
\dag}(\mathbf{p},E_{\mathbf{p}})\psi(\mathbf{r},t)e^{-i\mathbf{p}\mathbf{r}}\,\,.
\end{eqnarray}
Note that here the operators are explicitly time-dependent. This
corresponds to the Heisenberg picture. For mixed neutrinos, the
representation in Fourier space cannot diagonalize the energy
dependence and the momentum dependence simultaneously.

The Dirac spinors $u(\mathbf{p},E_{\mathbf{p}})$ and
$v(\mathbf{p},E_{\mathbf{p}})$ refer to negative-helicity particles
and positive-helicity antiparticles, respectively.\\

We need to incorporate the appropriate spin-statistics properties
into the description. At equal times the Pauli exclusion principle
is implemented by anti-commutation relations of the creation and
annihilation operators in momentum space:
\begin{eqnarray}
\{a(\mathbf{p},t),a^{\dag}(\mathbf{p'},t)\}=\{b(\mathbf{p},t),b^{\dag}(\mathbf{p'},t)\}
=(2\pi)^{3}\delta^{3}(\mathbf{p}-\mathbf{p'})\,\,.
\end{eqnarray}
One can now write the Wigner phase-space density operator for Dirac
neutrinos in terms of the creation and annihilation operators in
momentum space as
\begin{eqnarray}
\rho(\mathbf{r},\mathbf{p},t)=
\int\frac{d^{3}\mathbf{p'}}{(2\pi)^{3}}e^{-i\mathbf{p'}\mathbf{\mathbf{r}}}
a^{\dag}(\mathbf{p}-\frac{1}{2}\mathbf{p'},t)a(\mathbf{p}+\frac{1}{2}\mathbf{p'},t)\,\,.
\label{wignerop}
\end{eqnarray}
For antineutrinos we have the expression
\begin{eqnarray}
\tilde{\rho}(\mathbf{r},\mathbf{p},t)=
\int\frac{d^{3}\mathbf{p'}}{(2\pi)^{3}}e^{-i\mathbf{p'}\mathbf{\mathbf{r}}}
b^{\dag}(\mathbf{p}-\frac{1}{2}\mathbf{p'},t)b(\mathbf{p}+\frac{1}{2}\mathbf{p'},t)\,\,.
\end{eqnarray}
The same Wigner phase-space density operator can be formulated in
terms of the creation and annihilation operators in position space
as long as the integration variable is changed:
\begin{eqnarray}
\rho(\mathbf{r},\mathbf{p},t)=
\int\frac{d^{3}\mathbf{r'}}{(2\pi\hbar)^{3}}e^{-i\mathbf{p}\mathbf{\mathbf{r'}}}
\psi^{\dag}(\mathbf{r}-\frac{1}{2}\mathbf{r'},t)\psi(\mathbf{r}+\frac{1}{2}\mathbf{r'},t)\,\,,
\label{wignerop2}
\end{eqnarray}
and similarly for antineutrinos
\begin{eqnarray}
\tilde{\rho}(\mathbf{r},\mathbf{p},t)=
\int\frac{d^{3}\mathbf{r'}}{(2\pi\hbar)^{3}}e^{-i\mathbf{p}\mathbf{\mathbf{r'}}}
\tilde{\psi}^{\dag}(\mathbf{r}-\frac{1}{2}\mathbf{r'},t)\tilde{\psi}(\mathbf{r}+\frac{1}{2}\mathbf{r'},t)\,\,.
\end{eqnarray}
\\

The Wigner phase-space density operator possesses the sought-after
properties of a quantum-mechanical generalization of the classical
phase-space density. To see this, we integrate eq. (\ref{wignerop2})
with respect to $\mathbf{p}$ and use the representation of the
delta-function in Fourier space to write
\begin{eqnarray}
\int
d^{3}\mathbf{p}\rho(\mathbf{r},\mathbf{p},t)=\left|\psi\left(\mathbf{r},t\right)\right|^{2}\,\,,
\end{eqnarray}
which resembles the correct observable probability. When eq.
(\ref{wignerop2}) is integrated with respect to $\mathbf{r}$ one has
\begin{eqnarray}
\int d^{3}\mathbf{r}\rho(\mathbf{r},\mathbf{p},t)=\left|\int
d^{3}\mathbf{r}\psi(\mathbf{r},t)e^{-i\mathbf{p}\mathbf{r}}\right|^{2}=
\left|\psi\left(\mathbf{p},t\right)\right|^{2}\,\,,
\end{eqnarray}
which is again the correct expression. This can be achieved by a
shift of variables:
\begin{eqnarray}
\mathbf{r}-\frac{1}{2}\mathbf{r}'&=&\mathbf{v}\nonumber\\
\mathbf{r}+\frac{1}{2}\mathbf{r}'&=&\mathbf{u}\,\,,
\end{eqnarray}
and a subsequent relabeling of the integration variable. The above
derivation was done for fermions. The procedure is identical for
bosons however. Then, we impose commutation relations instead of the
anti-commutation relations for the operators. Furthermore no
restrictions to the mass of the mixed particle of
interest are made.\\

In the general case of mixed particles (this includes oscillating
fermions such as neutrinos but also mixing bosons such as
photon-axion conversion systems), the creation and annihilation
operators naturally yield mixed particles. To project these
composite objects onto their``irreducible'' subspaces we take the
matrix elements in the appropriate basis. We simultaneously perform
an ensemble average. We show that this transforms the Wigner
phase-space operator into its matrix elements. The matrix elements
are of great use since the diagonal elements are quasi-classical
phase-space densities. In the off-diagonal of this matrix we have
macroscopic overlap functions. One is now calculating numbers and
has not to worry about the time evolution of operators over Fock
space. Instead the value of a number changes with time.\\

In principle, these matrix elements are very similar to the density
matrix or statistical operator of a quantum-mechanical system as for
instance for the spin density matrix of particle beams.

The quantum number spin can formally be brought on the same page as
the quantum number flavor (Friedland \& Lunardini 2003). The
description of both systems respects the SU(N) symmetry for a
N-composite problem. The form of the interaction is alike and has
been useful as a calculational analogy.\\

In the following section \ref{sec:matrixelements}, we demonstrate
how the Wigner phase-space operator yields quasi-classical
phase-space densities.
\newpage
\section{Matrix elements}
\label{sec:matrixelements}
To decompose the Wigner phase-space density operator into
quasi-classical phase-space densities, scalar functions with
seven-dimensional domain, we need to calculate its matrix elements.
For clarity, we set the dimensions of flavor space to two. We
average over an ensemble consisting of two mixing particle species
and take the matrix elements in the number-density basis of Fock
space. This results in the expression
\begin{eqnarray} \mathcal{F}=\langle n_{i}|\rho|n_{j}\rangle &=&\left(
      \begin{matrix}
      f_{\nu_{e}}     &  f_{e\mu}     \\
      f^{\ast}_{e\mu} &  f_{\nu_{\mu}}\\
      \end{matrix}
\right)\,\,,\label{matrixelements}
\end{eqnarray}
where the indices $i$, $j$ run over $e$, $\mu$, and $^\ast$ means
complex conjugation\footnote{Following (de Groot, van Leeuwen \& van
Weert 1980), one can define the covariant energy-momentum densities
\begin{eqnarray}
T^{\mu\nu}(x)=c\int\frac{d^{3}p}{p^{0}}p^{\mu}p^{\nu}\text{tr}\,\,\mathcal{F}
\label{tensor}
\end{eqnarray}
and the conserved neutrino current densities
\begin{eqnarray}
J^{\mu}(x)=c\int\frac{d^{3}p}{p^{0}}p^{\mu}\text{tr}\left[\mathcal{F}-\tilde{\mathcal{F}}\right]\,\,,
\label{current}
\end{eqnarray}
wherein, by conservation of matter,
\begin{eqnarray}
\partial_{\mu}J^{\mu}=0.
\end{eqnarray}
$x^{\mu}=(ct,\mathbf{x})$ denotes a position four-vector and
$p^{\mu}=(E/c,\mathbf{p})$ the four-momentum respectively. With the
aid of these entities, eq. (\ref{current}) and eq. (\ref{tensor})
which describe a quasi-classical field, one can apply the framework
of classical field theory to investigate for instance the symmetries
and conserved quantities of the neutrino ensemble.}.
For the off-diagonal macroscopic overlap functions we write
\begin{eqnarray}
f_{e\mu}&=&\langle n_{\nu
_{e}}|\mathcal{\rho}(\mathbf{r},\mathbf{p},t)|n_{\nu_{\mu}}\rangle\nonumber\\
&=&\int\frac{d^{3}\mathbf{p'}}{(2\pi)^{3}}e^{i\mathbf{p'}\mathbf{\mathbf{r}}}\left\langle
n_{\nu_{e}}\right|
a^{\dag}(\mathbf{p}-\frac{1}{2}\mathbf{p'},t)a(\mathbf{p}+\frac{1}{2}\mathbf{p'},t)
\left|n_{\nu_{\mu}}\right\rangle\,\,, \label{flavoroverlap}
\end{eqnarray}
and for the $\nu_{e}$ neutrino phase-space density we have the
expression:
\begin{eqnarray}
f_{\nu_{e}}&=&\langle n_{\nu
_{e}}|\mathcal{\rho}(\mathbf{r},\mathbf{p},t)|n_{\nu_{e}}\rangle\nonumber\\
&=&\int\frac{d^{3}\mathbf{p'}}{(2\pi)^{3}}e^{i\mathbf{p'}\mathbf{\mathbf{r}}}\left\langle
n_{\nu_{e}}\right|
a^{\dag}(\mathbf{p}-\frac{1}{2}\mathbf{p'},t)a(\mathbf{p}+\frac{1}{2}\mathbf{p'},t)
\left|n_{\nu_{e}}\right\rangle\,\,,
\end{eqnarray}
and identically for the $\nu_{\mu}'s$.\\

Heuristically, we can put this process of simultaneous projection
and averaging as follows. The integration process is symmetric with
respect to $\mathbf{p'}$. Thus, each of the
$a^{\dag}(\mathbf{p}-\frac{1}{2}\mathbf{p'},t)a(\mathbf{p}+\frac{1}{2}\mathbf{p'},t)$
hits every momentum value of the $n_{\nu_{e}}$. Regarded this way,
we can anti-commute $a^{\dag}(\mathbf{p}-\frac{1}{2}\mathbf{p'},t)$
and $a(\mathbf{p}+\frac{1}{2}\mathbf{p'},t)$ in momentum space until
they are adjacent and possess identical arguments. At that instant
they act on their eigenstates and yield numbers as their
eigenvalues. Subsequently, we are left with real-valued functions
possessing seven-dimensional domain; quasi-classical phase-space
densities $f_{\nu_{i}}$.

For completeness, we list the matrix elements for the associated
antiparticles:
\begin{eqnarray}\tilde{\mathcal{F}}=\langle \tilde{n}_{i}|\tilde{\rho}|\tilde{n}_{j}\rangle &=&\left(
      \begin{matrix}
      \tilde{f}_{\nu_{e}}         &  \tilde{f}_{e\mu} \\
      \tilde{f}^{\ast}_{e\mu} &  \tilde{f}_{\nu_{\mu}}  \\
      \end{matrix}
\right)\,\,.\label{antimatrixelements}
\end{eqnarray}
\\

Note that it is the hermiticity of $\rho$ and $\tilde{\rho}$ that
implies the off-diagonal terms of its matrix elements be
complex-conjugates of one another.\\

A generalization to more particle species is straightforward. The
diagonal terms are real-valued and denote quasi-classical
phase-space densities. The off-diagonal entries are complex-valued
macroscopic overlap functions. They represent the degree of
coherence or macroscopic overlap of the ensemble under
consideration. For completely decohered ensembles and for non-mixing
ensembles the off-diagonal entries vanish. More precisely,
$f_{e\mu}(\mathbf{r},\mathbf{p},t)$ for instance gives the number of
$\nu_{e}$'s that result from incoherently adding up the quantum
mechanical amplitudes from each otherwise $\nu_{\mu}$. This requires
that flavor-entangled many-body states can be neglected or are small
in number.
\section{Conclusions}
\begin{itemize}

\item The salient results of chapter \ref{wigner} are eqs.
(\ref{matrixelements}) and (\ref{antimatrixelements}).\\

\item We have abandoned the operator character of the mathematical
entities describing the neutrinos, eq. (\ref{wignerop}), and are now
in position to work with real-valued phase-space densities and their
complex-number-valued overlap correlates, eq.
(\ref{matrixelements}).\\

\item In the following, we construct dynamical equations for the quasi-classical phase-space
densities which are very similar to the equations of classical
transport theory (see for instance eq.
(\ref{radiationfield})).\\

\end{itemize}

\chapter[Generalized Boltzmann Equations]{Generalized Boltzmann Equations}
\label{quasi}
\section{$2\times2$ Mixing Hamiltonian}
Naively, we can derive a dynamical equation for the matrix elements
of the Wigner phase-space density operator by merging the Boltzmann
equation with the Heisenberg equation. A derivation from
first-principle quantum mechanics has been performed in many
different contexts (Landau 1957, Akhiezer \& Peletminskii 1981,
Rudszky 1990, Raffelt, Sigl \& Stodolsky 1993). We need to account
for the matrix character of the phase-space densities and, following
the nice papers of Raffelt \& Sigl 1993, generalize the ``Liouville
terms'' on the left-hand side by introducing anti-commutators
resulting in the expression
\begin{eqnarray}
\frac{\partial\mathcal{F}}{\partial
t}+\frac{1}{2}\left\{\mathbf{v},\frac{\partial\mathcal{F}}{\partial
\mathbf{r}}\right\}+\frac{1}{2}\left\{\dot{\mathbf{p}},\frac{\partial
\mathcal{F}}{\partial \mathbf{p}}\right\}
=-i\left[\Omega,\mathcal{F}\right]+\mathcal{C}\,\,.\nonumber\\
\label{heisboltz}
\end{eqnarray}

Here, the collision matrix is given by
\begin{eqnarray}
\mathcal{C}(\mathbf{r},\mathbf{p},t)=\left(\begin{matrix}
\mathcal{C}_{\nu_{e}}& \mathcal{C}_{\nu_{e}\rightarrow \nu_{\mu}}\\
\mathcal{C}_{\nu_{\mu}\rightarrow \nu_{e}} & \mathcal{C}_{\nu_{\mu}}
\end{matrix}
\right)\,\,,\label{coll}
\end{eqnarray}
where the off-diagonal elements describe flavor-changing collisions.
In the standard model, flavor number is conserved at every vertex.
So by restricting to these interactions one can in general set the
off-diagonal elements to zero.

Oscillation phenomenology is incorporated via the commutator on the
right-hand-side, the term that originated in the Heisenberg
equation, with the mixing Hamiltonian given by,
\begin{eqnarray}
\Omega(\mathbf{r},\mathbf{p})=\Omega_{\text{vac}}(\varepsilon)+\Omega_{\text{mat}}(\mathbf{r})+
\Omega_{\nu\nu}(\mathbf{r},\mathbf{p},t)-\tilde{\Omega}_{\tilde{\nu}\nu}(\mathbf{r},\mathbf{p},t)\,\,.
\label{mixhamilton}
\end{eqnarray}
In this cumulative expression, $\Omega_{\text{vac}}(\varepsilon)$ is
the usual vacuum contribution where mixing comes about through the
non-degeneracy of masses and the entailed distinction between flavor
or weak interaction eigenstates and the mass eigenstates.
$\Omega_{\text{mat}}(\mathbf{r})$ is the ordinary matter
contribution that most notably results the effective mass for the
$\nu_{e}$ neutrino in matter. $\Omega_{\nu\nu}$ is the non-diagonal
Hamiltonian that describes the neutrino self-interactions or
neutrino-neutrino scattering and $\tilde{\Omega}_{\nu\tilde{\nu}}$
is the non-diagonal neutrino-antineutrino
interaction Hamiltonian.\\

Making use of the antineutrino phase-space densities as given in eq.
(\ref{antimatrixelements}) the dynamical equation for the
anti-particles looks completely analogous,
\begin{eqnarray}
\frac{\partial\tilde{\mathcal{F}}}{\partial
t}+\frac{1}{2}\left\{\mathbf{v},\frac{\partial\tilde{\mathcal{F}}}{\partial
\mathbf{r}}\right\}+\frac{1}{2}\left\{\dot{\mathbf{p}},\frac{\partial
\tilde{\mathcal{F}}}{\partial \mathbf{p}}\right\}
=-i\left[\tilde{\Omega},\tilde{\mathcal{F}}\right]+\tilde{\mathcal{C}}\,\,,\nonumber\\
\label{antiheisboltz}
\end{eqnarray}
where apart from the obvious change in the collision terms and the
different phase-space densities (denoted by a tilde), we must
reverse sign for the coupling coefficients in $\Omega_{\text{mat}}$,
$\Omega_{\nu\nu}$ and $\tilde{\Omega}_{\tilde{\nu}\nu}$,
\begin{eqnarray}
\tilde{\Omega}(\mathbf{r},\mathbf{p})=\Omega_{\text{vac}}(\varepsilon)-\Omega_{\text{mat}}(\mathbf{r})
-\Omega_{\nu\nu}(\mathbf{p})+\tilde{\Omega}_{\tilde{\nu}\nu}(\mathbf{p})\,\,.
\label{antimixhamilton}
\end{eqnarray}
The vacuum mixing contribution is alike for neutrinos and its
antiparticle,
\begin{eqnarray}
\Omega_{\text{vac}}(\varepsilon)=\frac{\pi c}{L}\left(\begin{matrix}
-\cos 2 \theta & \sin 2\theta\\
\sin 2\theta & \cos 2 \theta
\end{matrix}
\right)\,\,,
\end{eqnarray}
where $\theta_{vac}$ is the mixing angle for vacuum neutrino
oscillations between $\nu_{e}$'s and $\nu_{\mu}$'s and $L$ is the
vacuum neutrino oscillation length:
\begin{eqnarray}
L=\frac{4\pi\hbar c\varepsilon}{\Delta m^{2}c^{4}}\,\,,
\end{eqnarray}
where $\varepsilon$ is the neutrino energy, $m_{1}$ and $m_{2}$ are
the masses of the neutrino mass eigenstates, and $\Delta
m^{2}=m_{2}^{2}-m_{1}^{2}$. The other variables have their standard
meanings.

The electron neutrino finds the most leptonic interaction partners
in the ambient matter. Therefore, we have the matrix:
\begin{eqnarray}
\Omega_{\text{mat}}(\mathbf{r})=\left(\begin{matrix}
2A & 0\\
0  & 0
\end{matrix}
\right)\,\,, \label{standardeffmass}
\end{eqnarray}
where $A$ is the dimensionless neutrino-matter interaction amplitude
(Mohapatra \& Pal 2003, Giunti 2004):
\begin{eqnarray}
A(\mathbf{r})=\frac{2\sqrt{2}G_{F}}{\hbar}n_{e}(\mathbf{r})\,\,,
\end{eqnarray}
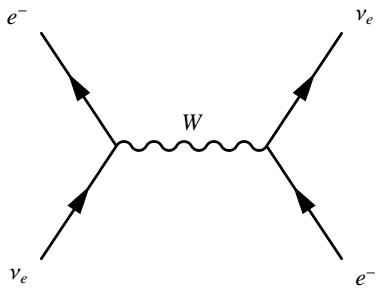
\begin{figure}
\begin{center}
\begin{fmffile}{chargedcurrent}
 \begin{fmfgraph*}(50,30)
  \fmfleft{l1,l2}
  \fmfright{r1,r2}
   \fmflabel{$\nu_{e}$}{l1}
   \fmflabel{$e^{-}$}{l2}
   \fmflabel{$e^{-}$}{r1}
   \fmflabel{$\nu_{e}$}{r2}
  \fmf{fermion}{l1,v1,l2}
  \fmf{fermion}{r1,v2,r2}
  \fmf{boson,label=$W$}{v1,v2}
\end{fmfgraph*}
\end{fmffile}
\\[10mm]
\caption{Charged-current scattering reaction leading to the
effective mass of the $\nu_{e}$ in an electron background. $W$ is
the charged current gauge boson mediating the weak force.}
\label{chargedcurrent}
\end{center}
\end{figure}
where $G_{F}$ denotes Fermi's constant, and $n_{e}$ the electron
number density. The interaction amplitude $A$, through $n_e$, is a
function of spatial position. For antineutrinos the sign of $A$ is
reversed
as indicated in eq. (\ref{antimixhamilton}).\\

Diagrammatically, the charged-current reaction contributing to the
effective mass of the $\nu_{e}$ is given in Fig.
(\ref{chargedcurrent}).

Clearly, the $\nu_{\mu}$'s and $\nu_{\tau}$'s find orders of
magnitude less scattering partners in the ambient matter. Arguments
can be given that the neutral-current interactions involving the $Z$
boson are the same for all flavors and thus are not included in the
mixing Hamiltonian; here we are only interested in differences
between the flavor species (Mohapatra \& Pal 2003). For an
illustrative scenario where this neutral-current scattering can be
important see section \ref{sec:curvedmixing} and Fig.
\ref{neutralcurrent}.
\section{Neutrino self-interactions}
The neutrino evolution in a neutrino background is nontrivial and
continues to raise considerable discussions in the literature
(Pantaleone 1992 A, Pantaleone 1995, Qian \& Fuller 1995, Friedland
\& Lunardini 2003, Bell, Rawlinson \& Sawyer 2003, Boyanovsky \& Ho
2004, Sawyer 2004, and Sawyer 2005). Here we sketch the
considerations that led to $\Omega_{\nu\nu}$ and work with the most
established version that can be found in the literature.\\

New is, that we match the wavefunctions to the quasi-classical
phase-space densities employed in our ensemble-averaged approach. In
this work, we do not engage in efforts about a possible speed-up of
flavor equilibration as proposed by Sawyer, \etal . We retain the
one-particle description neglecting any effects arising from
coherent many-body state formation or flavor entanglement (Friedland
\& Lunardini 2003). The exact expression for the neutral-current
neutrino-neutrino interaction Hamiltonian for a test neutrino with
initial momentum $\mathbf{p}$ and the background neutrino with
initial momentum $\mathbf{q}$ is given by (Raffelt \& Sigl 1993):
\begin{eqnarray}
H^{\mathbf{pq}}_{\nu\nu}=\frac{1}{2}\left(\frac{G_{F}}{2\cos\theta_{W}}\right)^{2}\int
d\mathbf{p}'\int d \mathbf{q}'
\left(2\pi\right)^{3}\delta^{(3)}\left(\mathbf{p}+\mathbf{q}-\mathbf{p}'-\mathbf{q}'\right)
\bar{\psi}_{\mathbf{q}}\gamma^{\mu}\psi_{\mathbf{q}'}D^{Z}_{\mu\nu}
\bar{\psi}_{\mathbf{p}}\gamma^{\nu}\psi_{\mathbf{p}'}\,\,,
\nonumber\\
\end{eqnarray}
where the $\psi$'s are the Dirac neutrino spinors as specified in
eq. (\ref{nuop}), $\gamma^{\mu}$ are the standard Dirac matrices as
given in eq. (\ref{diracalgebra}), $\cos\theta_{W}$ is the Weinberg
angle, $g$ is the weak coupling constant given by
\begin{eqnarray}
g=\left(4\sqrt{2}G_{F}\right)^{1/2}m_{Z}\,\,,
\end{eqnarray}
where $G_F$ is Fermi's constant. $D^{Z}_{\mu\nu}$ is the Z-boson
propagator
\begin{eqnarray}
D^{Z}_{\mu\nu}(p)=\left(\eta_{\mu\nu}-\frac{p^{\mu}p^{\nu}}{m_{Z}^{2}}\right)
\frac{1}{m^{2}_{Z}-p^{2}}\,\,.
\end{eqnarray}
Force mediating gauge bosons that propagate from one neutrino at a
given spacetime point to another neutrino at another spacetime point
is a non-local effect. This can be readily discarded since in
supernova cores and in most other situations found in nature one is
typically interested in neutrinos with energies of the order of MeV.
This is well below the mass or energy scale of the Z-boson of order
100 GeV. Therefore, we neglect the non-local gauge boson effects and
work with the low-energy local four-fermion coupling. This effective
Hamiltonian must satisfy SU(N) flavor symmetry for a system of N
flavors (Pantaleone 1992 A) and we write the following terms:
\begin{eqnarray}
\Omega^{\mathbf{pq}}_{\nu\nu}=\frac{G_{F}}{\sqrt{2}}\left(\sum_{i}
\bar{\psi}^{i}_{\mathbf{q}}\gamma^{\mu}\psi^{i}_{\mathbf{q}}\right)
\left(\sum_{j}\bar{\psi}^{j}_{\mathbf{q}}\gamma_{\mu}\psi^{j}_{\mathbf{q}}\right)\,\,,
\end{eqnarray}
where the sum is over all neutrino flavors. In this form the SU(N)
flavor symmetry is manifest. For a two-flavor system consisting of
$\nu_{e}$'s and $\nu_{\mu}$'s  one can rewrite this Hamiltonian to
accentuate its off-diagonal character (Pantaleone 1992 B, Friedland
\& Lunardini 2003):
\begin{eqnarray}
\Omega^{\mathbf{pq}}_{\nu\nu}&=&\beta\left(1-\cos\theta^{\mathbf{pq}}\right)
\left[\Big|\psi_{\nu_{e}}^{\mathbf{q}}\Big|^{2}
+\Big|\psi_{\nu_{\mu}}^{\mathbf{q}}\Big|^{2}+ \left(\begin{matrix}
\Big|\psi_{\nu_{e}}^{\mathbf{q}}\Big|^{2} & \psi_{\nu_{e}}^{\mathbf{q}}\psi_{\nu_{\mu}}^{\mathbf{q}\ast}\\
\psi_{\nu_{e}}^{\mathbf{q}\ast}\psi_{\nu_{\mu}}^{\mathbf{q}}
&\left|\psi_{\nu_{\mu}}^{\mathbf{q}}\right|^{2}\end{matrix}
\right)\right]\,\,, \label{nunu}
\end{eqnarray}
where the coupling coefficient prefers diametral scattering by
inclusion of the angle $\theta$ between the test neutrino with
momentum $\mathbf{p}$ and the background neutrino with momentum
$\mathbf{q}$.
\newpage
\begin{figure}
\begin{center}
\begin{fmffile}{diagself1}
 \begin{fmfgraph*}(50,30)
  \fmfleft{l1,l2}
  \fmfright{r1,r2}
   \fmflabel{$\nu_{e}$}{l1}
   \fmflabel{$\nu_{e}'$}{l2}
   \fmflabel{$\nu_{e}'$}{r1}
   \fmflabel{$\nu_{e}$}{r2}
  \fmf{fermion}{l1,v1,l2}
  \fmf{fermion}{r1,v2,r2}
  \fmf{boson,label=$Z$}{v1,v2}
\end{fmfgraph*}
\end{fmffile}
\hspace{10mm}
\begin{fmffile}{diagself2}
 \begin{fmfgraph*}(50,30)
  \fmfleft{l1,l2}
  \fmfright{r1,r2}
   \fmflabel{$\nu_{\mu}$}{l1}
   \fmflabel{$\nu_{\mu}'$}{l2}
   \fmflabel{$\nu_{\mu}'$}{r1}
   \fmflabel{$\nu_{\mu}$}{r2}
  \fmf{fermion}{l1,v1,l2}
  \fmf{fermion}{r1,v2,r2}
  \fmf{boson,label=$Z$}{v1,v2}
\end{fmfgraph*}
\end{fmffile}\\[10mm]
\caption{Diagonal contribution to the neutrino-selfinteractions:
$\nu_{e}-\nu_{e}$ and $\nu_{\mu}-\nu_{\mu}$ scattering.}
\label{diagnunu}
\end{center}
\end{figure}
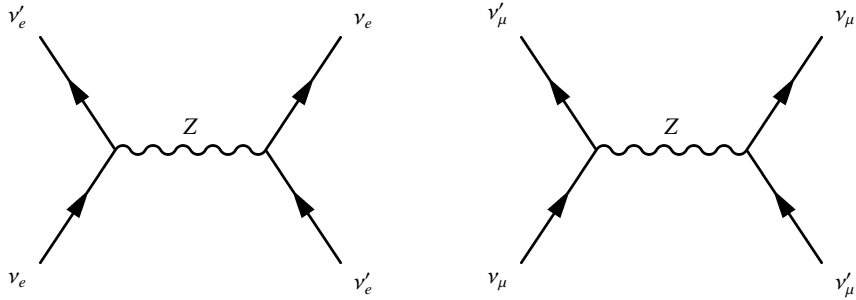
\vspace{10mm}
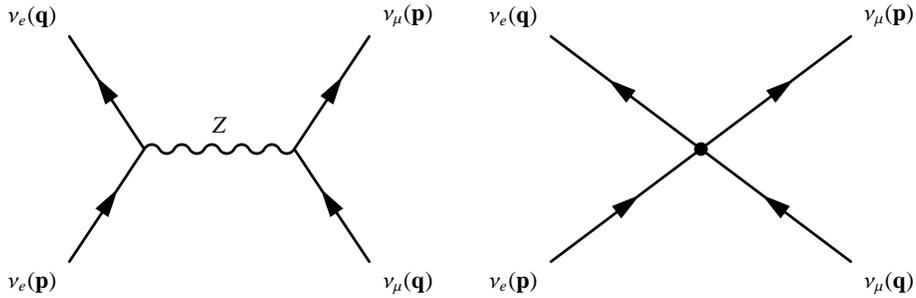
\begin{figure}
\begin{center}
\begin{fmffile}{offdiagnunu1}
 \begin{fmfgraph*}(50,30)
  \fmfleft{l1,l2}
  \fmfright{r1,r2}
   \fmflabel{$\nu_{e}(\mathbf{p})$}{l1}
   \fmflabel{$\nu_{e}(\mathbf{q})$}{l2}
   \fmflabel{$\nu_{\mu}(\mathbf{q})$}{r1}
   \fmflabel{$\nu_{\mu}(\mathbf{p})$}{r2}
  \fmf{fermion}{l1,v1,l2}
  \fmf{fermion}{r1,v2,r2}
  \fmf{boson,label=$Z$}{v1,v2}
\end{fmfgraph*}
\end{fmffile}
\hspace{10mm}
\begin{fmffile}{offdiagnunu2}
 \begin{fmfgraph*}(50,30)
  \fmfleft{l1,l2}
  \fmfright{r1,r2}
   \fmflabel{$\nu_{e}(\mathbf{p})$}{l1}
   \fmflabel{$\nu_{e}(\mathbf{q})$}{l2}
   \fmflabel{$\nu_{\mu}(\mathbf{q})$}{r1}
   \fmflabel{$\nu_{\mu}(\mathbf{p})$}{r2}
  \fmf{fermion}{l1,v,l2}
  \fmf{fermion}{r1,v,r2}
  \fmfdot{v}
  %\fmf{boson,label=$Z$}{v1,v2}
\end{fmfgraph*}
\end{fmffile}\\[10mm]
\caption{Off-diagonal contribution to the
neutrino-self-interactions. The $Z$-boson carries the difference in
momenta. The graph on the right-hand-side is the low-energy local
four-fermion interaction.} \label{offdiagnunu}
\end{center}
\end{figure}
The coupling strength $\beta$ and the normalization of the
background neutrino fields are determined by the formulae:
\begin{eqnarray}
\int dV
\Big|\psi_{\nu_{e}}^{\mathbf{q}}\Big|^{2}+\Big|\psi_{\nu_{\mu}}^{\mathbf{q}}\Big|^{2}&=&1\nonumber\\[3mm]
\beta&=&\frac{\sqrt{2}G_{F}}{\hbar}\,\,.
\end{eqnarray}
\\

In Fig. \ref{diagnunu}, the two diagonal contributions to the
$\nu-\nu$ interaction are shown. The time axis is vertical. In Fig.
\ref{offdiagnunu}, the off-diagonal contributions to the $\nu-\nu$
interaction are depicted. Here the point is that in forward
direction the momenta of the two scattering partners are exchanged.
Since the momentum eigenstates are the propagation eigenstates in
the ambient matter, this is considered as a flavor-exchanging
reaction.\\

We now match the the $\nu-\nu$ Hamiltonian from the wavefunction
formalism to our ensemble-averaged description. To this end, we make
use of our quasi-classical phase-space densities and the macroscopic
overlap functions of eq. (\ref{matrixelements}). This involves
integration over all the momenta $\mathbf{q}$ of the ensemble (note
that the neutrino ensemble also constitutes the background).\\

The $\nu$--$\nu$ mixing Hamiltonian for test neutrinos with momentum
$\mathbf{p}$ in ensemble-averaged form is denoted by
\begin{eqnarray}
\Omega_{\nu\nu}(\mathbf{p},\mathbf{r},t)=\left(\begin{matrix}
B_{\nu_{e}} & B_{e\mu}\\
B^{\ast}_{e\mu} & B_{\nu_{\mu}}\end{matrix} \right)\,\,, \label{B's}
\end{eqnarray}
where, while throwing away the overall phase term in eq.
(\ref{nunu}), which is proportional to the identity matrix, we write
for the diagonal elements:
\begin{eqnarray}
B_{\nu_{e}}(\mathbf{p},\mathbf{r},t)&=&\beta\int
d^{3}\mathbf{q}\left(1-\cos\theta^{\mathbf{pq}}\right)f_{\nu_{e}}(\mathbf{q},\mathbf{r},t)\nonumber\\
B_{\nu_{\mu}}(\mathbf{p},\mathbf{r},t)&=&\beta\int
d^{3}\mathbf{q}\left(1-\cos\theta^{\mathbf{pq}}\right)f_{\nu_{\mu}}(\mathbf{q},\mathbf{r},t)\,\,,
\end{eqnarray}
where the momentum integration goes over all momenta in the
ensemble. For the off-diagonal elements we write in our notation, in
terms of the macroscopic overlap functions,
\begin{eqnarray}
B_{e\mu}(\mathbf{p},\mathbf{r},t)&=&\beta\int
d^{3}\mathbf{q}\left(1-\cos\theta^{\mathbf{pq}}\right)f_{e\mu}(\mathbf{q},\mathbf{r},t)\nonumber\\
B^{\ast}_{e\mu}(\mathbf{p},\mathbf{r},t)&=&\beta\int
d^{3}\mathbf{q}\left(1-\cos\theta^{\mathbf{pq}}\right)f^{\ast}_{e\mu}(\mathbf{q},\mathbf{r},t)\,\,.
\end{eqnarray}
For the antineutrinos in the ensemble and the ambient matter,
\begin{eqnarray}
\tilde{\Omega}_{\tilde{\nu}\nu}(\mathbf{p},\mathbf{r},t)=\left(\begin{matrix}
\tilde{B}_{\nu_{e}} & \tilde{B}_{e\mu}\\
\tilde{B}^{\ast}_{e\mu} &\tilde{B}_{\nu_{\mu}}\end{matrix}
\right)\,\,.
\end{eqnarray}
And in complete analogy, we write for the interactions of the test
neutrino with momentum $\mathbf{p}$ with the antineutrinos,
\begin{eqnarray}
\tilde{B}_{\nu_{e}}(\mathbf{p},\mathbf{r},t)&=&\beta\int
d^{3}\mathbf{q}\left(1-\cos\theta^{\mathbf{pq}}\right)\tilde{f}_{\nu_{e}}(\mathbf{q},\mathbf{r},t)\nonumber\\
\tilde{B}_{\nu_{\mu}}(\mathbf{p},\mathbf{r},t)&=&\beta\int
d^{3}\mathbf{q}\left(1-\cos\theta^{\mathbf{pq}}\right)\tilde{f}_{\nu_{\mu}}(\mathbf{q},\mathbf{r},t)\nonumber\\
\tilde{B}_{e\mu}(\mathbf{p},\mathbf{r},t)&=&\beta\int
d^{3}\mathbf{q}\left(1-\cos\theta^{\mathbf{pq}}\right)\tilde{f}_{e\mu}(\mathbf{q},\mathbf{r},t)\nonumber\\
\tilde{B}^{\ast}_{e\mu}(\mathbf{p},\mathbf{r},t)&=&\beta\int
d^{3}\mathbf{q}\left(1-\cos\theta^{\mathbf{pq}}\right)\tilde{f}^{\ast}_{e\mu}(\mathbf{q},\mathbf{r},t)\,\,.
\label{nuantinu}
\end{eqnarray}
The coupling coefficient in front of the integrals has to be
implemented in the equations for neutrinos with reversed sign. For a
similar way of parameterizing the neutrino--neutrino Hamiltonian see
the nice paper by Fuller \& Qian 1995.
\section{Kinetic equations}
We are interested in an analysis of the macroscopic, long-range
transport behavior of eq. (\ref{heisboltz}) and execute the
following simplifying procedure.
\begin{itemize}

\item We do not take into account the effects of small-scale external
homogeneities on the mixing in the ensemble (Sawyer 1990).\\

\item We average out any small-scale correlations and do not include any
other off-diagonal contributions from the complete Hamiltonian.\\

\item We neglect off-diagonal contributions from the left-hand-side
of eq. (\ref{heisboltz}).\\

\item We componentize eq. (\ref{heisboltz}) and include all the
contributions from the mixing Hamiltonian in eq.
(\ref{mixhamilton}).\\
\end{itemize}

To this end it turns out to be useful to define the real part of the
off-diagonal macroscopic overlap in eq. (\ref{matrixelements}) as
\begin{eqnarray}
f_{r}=\frac{1}{2}\left(f_{e\mu}+f^{\ast}_{e\mu}\right)\,\,,
\end{eqnarray}
and the corresponding imaginary part as
\begin{eqnarray}
f_{i}=\frac{1}{2i}\left(f_{e\mu}-f^{\ast}_{e\mu}\right)\,\,,\nonumber\\
\label{offdia}
\end{eqnarray}
respectively. For antineutrinos one proceeds analogously. For the
real part one has
\begin{eqnarray}
\tilde{f_{r}}=\frac{1}{2}\left(\tilde{f}_{e\mu}+\tilde{f}^{\ast}_{e\mu}\right)\,\,,
\end{eqnarray}
and for the corresponding imaginary part there is
\begin{eqnarray}
\tilde{f_{i}}=\frac{1}{2i}\left(\tilde{f}_{e\mu}-\tilde{f}^{\ast}_{e\mu}\right)\,\,.\nonumber\\
\label{antioffdia}
\end{eqnarray}

With this choice of parameters, everything becomes real-valued and
the $i's$ are eliminated from the equations. In the standard weak
interaction basis or flavor basis the collision terms are diagonal
and we will not consider flavor-changing reactions here. Henceforth
we apply the diagonal collision matrix,
\begin{eqnarray}
\mathcal{C}=\left(\begin{matrix}
\mathcal{C}_{\nu_{e}}& 0\\
0& \mathcal{C}_{\nu_{\mu}}
\end{matrix}
\right).
\end{eqnarray}
For antineutrinos the collision terms differ and are denoted by
\begin{eqnarray}
\tilde{\mathcal{C}}=\left(\begin{matrix}
\tilde{\mathcal{C}}_{\nu_{e}}& 0\\
0& \tilde{\mathcal{C}}_{\nu_{\mu}}
\end{matrix}
\right).
\end{eqnarray}
Having made these definitions, we are in a position to componentize
eq. (\ref{heisboltz}) and eq. (\ref{antiheisboltz}). We work with
the general mixing Hamiltonian eq. (\ref{mixhamilton}) and eq.
(\ref{antimixhamilton}). For two mixing neutrino species, here
denoted by $\nu_{e}$ and $\nu_{\mu}$, interacting with a background
medium the generalized Boltzmann equations in their most generic
form are:

\begin{eqnarray}
\frac{\partial f_{\nu_{e}}}{\partial
t}+\mathbf{v}\cdot\frac{\partial f_{\nu_{e}}}{\partial\mathbf{r}}+
\mathbf{\dot{p}}\cdot\frac{\partial
f_{\nu_{e}}}{\partial\mathbf{p}}&=& -f_{i}\left(\frac{2\pi
c}{L}\sin2\theta +2\beta\int
\left(1-\cos\theta^{\mathbf{pq}}\right)\left(f_{r}+
\tilde{f_{r}}\right)d^{3}\mathbf{q}\right)+\nonumber\\
&&2\beta f_{r}\int\left(1-\cos\theta^{\mathbf{pq}}\right)
\left(f_{i}+\tilde{f_{i}}\right)d^{3}\mathbf{q}
+\mathcal{C}_{\nu_{e}}\nonumber\\
\frac{\partial f_{\nu_{\mu}}}{\partial
t}+\mathbf{v}\cdot\frac{\partial f_{\nu_{\mu}}}{\partial\mathbf{r}}+
\mathbf{\dot{p}}\cdot\frac{\partial
f_{\nu_{\mu}}}{\partial\mathbf{p}}&=& f_{i}\left(\frac{2\pi
c}{L}\sin2\theta +2\beta\int
\left(1-\cos\theta^{\mathbf{pq}}\right)\left(f_{r}+
\tilde{f_{r}}\right)d^{3}\mathbf{q}\right)-\nonumber\\
&&2\beta f_{r}\int\left(1-\cos\theta^{\mathbf{pq}}\right)
\left(f_{i}+\tilde{f_{i}}\right)d^{3}\mathbf{q}
+\mathcal{C}_{\nu_{\mu}}\nonumber\\
\frac{\partial f_{r}}{\partial t}+\mathbf{v}\cdot\frac{\partial
f_{r}}{\partial \mathbf{r}}+\mathbf{\dot{p}} \cdot\frac{\partial
f_{r}}{\partial \mathbf{p}}&=& f_{i}\Big[\frac{2\pi
c}{L}\left(A-\cos2\theta\right)+\nonumber\\
&&\beta\int\left(1-\cos^{\mathbf{pq}}\right)
\left(f_{\nu_{e}}-\tilde{f}_{\nu_{e}}-f_{\nu_{\mu}}+\tilde{f}_{\nu_{\mu}}\right)d^{3}\mathbf{q}\Big]+\nonumber\\
&&\left(f_{\nu_{e}}-f_{\nu_{\mu}}\right)\beta\int\left(1-\cos\theta^{\mathbf{pq}}\right)
\left(\tilde{f}_{i}-f_{i}\right)d^{3}\mathbf{q}\nonumber\\
\frac{\partial f_{i}}{\partial t}+\mathbf{v}\cdot\frac{\partial
f_{i}}{\partial \mathbf{r}}+\mathbf{\dot{p}} \cdot\frac{\partial
f_{i}}{\partial
\mathbf{p}}&=&\left(f_{\nu_{e}}-f_{\nu_{\mu}}\right)\left[\frac{\pi
c}{L}\sin2\theta+\beta\int\left(1-\cos^{\mathbf{pq}}\right)
\left(f_{r}-\tilde{f}_{r}\right)d^{3}\mathbf{q}\right]-\nonumber\\
&&f_{r}\Big[\frac{2\pi
c}{L}\left(A-\cos2\theta\right)+\nonumber\\
&&\beta\int\left(1-\cos^{\mathbf{pq}}\right)
\left(f_{\nu_{e}}-\tilde{f}_{\nu_{e}}-f_{\nu_{\mu}}+\tilde{f}_{\nu_{\mu}}\right)d^{3}\mathbf{q}\Big]\,\,.
\label{genericboltz}
\end{eqnarray}
In a very similar fashion we can write the corresponding
antiparticle analog. We need to interchange and add tildes,
sign-reverse the matter interaction amplitude $A$, and complete our
set of equations:
\begin{eqnarray}
\frac{\partial \tilde{f}_{\nu_{e}}}{\partial
t}+\mathbf{v}\cdot\frac{\partial
\tilde{f}_{\nu_{e}}}{\partial\mathbf{r}}+
\mathbf{\dot{p}}\cdot\frac{\partial
\tilde{f}_{\nu_{e}}}{\partial\mathbf{p}}&=&
-\tilde{f}_{i}\left(\frac{2\pi c}{L}\sin2\theta +2\beta\int
\left(1-\cos\theta^{\mathbf{pq}}\right)\left(\tilde{f}_{r}+
f_{r}\right)d^{3}\mathbf{q}\right)+\nonumber\\
&&2\beta \tilde{f}_{r}\int\left(1-\cos\theta^{\mathbf{pq}}\right)
\left(\tilde{f}_{i}+f_{i}\right)d^{3}\mathbf{q}
+\tilde{\mathcal{C}}_{\nu_{e}}\nonumber\\
\frac{\partial \tilde{f}_{\nu_{\mu}}}{\partial
t}+\mathbf{v}\cdot\frac{\partial
\tilde{f}_{\nu_{\mu}}}{\partial\mathbf{r}}+
\mathbf{\dot{p}}\cdot\frac{\partial
\tilde{f}_{\nu_{\mu}}}{\partial\mathbf{p}}&=&
\tilde{f}_{i}\left(\frac{2\pi c}{L}\sin2\theta +2\beta\int
\left(1-\cos\theta^{\mathbf{pq}}\right)\left(\tilde{f}_{r}+
f_{r}\right)d^{3}\mathbf{q}\right)-\nonumber\\
&&2\beta \tilde{f}_{r}\int\left(1-\cos\theta^{\mathbf{pq}}\right)
\left(\tilde{f}_{i}+f_{i}\right)d^{3}\mathbf{q}
+\tilde{\mathcal{C}}_{\nu_{\mu}}\nonumber\\
\frac{\partial \tilde{f}_{r}}{\partial
t}+\mathbf{v}\cdot\frac{\partial \tilde{f}_{r}}{\partial
\mathbf{r}}+\mathbf{\dot{p}} \cdot\frac{\partial
\tilde{f}_{r}}{\partial \mathbf{p}}&=& \tilde{f}_{i}\Big[\frac{2\pi
c}{L}\left(-A-\cos2\theta\right)+\nonumber\\
&&\beta\int\left(1-\cos^{\mathbf{pq}}\right)
\left(\tilde{f}_{\nu_{e}}-f_{\nu_{e}}-\tilde{f}_{\nu_{\mu}}+f_{\nu_{\mu}}\right)d^{3}\mathbf{q}\Big]+\nonumber\\
&&\left(\tilde{f}_{\nu_{e}}-\tilde{f}_{\nu_{\mu}}\right)\beta\int\left(1-\cos\theta^{\mathbf{pq}}\right)
\left(f_{i}-\tilde{f}_{i}\right)d^{3}\mathbf{q}\nonumber\\
\frac{\partial \tilde{f}_{i}}{\partial
t}+\mathbf{v}\cdot\frac{\partial \tilde{f}_{i}}{\partial
\mathbf{r}}+\mathbf{\dot{p}} \cdot\frac{\partial
\tilde{f}_{i}}{\partial
\mathbf{p}}&=&\left(\tilde{f}_{\nu_{e}}-\tilde{f}_{\nu_{\mu}}\right)\left[\frac{\pi
c}{L}\sin2\theta+\beta\int\left(1-\cos^{\mathbf{pq}}\right)
\left(\tilde{f}_{r}-f_{r}\right)d^{3}\mathbf{q}\right]-\nonumber\\
&&\tilde{f}_{r}\Big[\frac{2\pi
c}{L}\left(-A-\cos2\theta\right)+\nonumber\\
&&\beta\int\left(1-\cos^{\mathbf{pq}}\right)
\left(\tilde{f}_{\nu_{e}}-f_{\nu_{e}}-\tilde{f}_{\nu_{\mu}}+f_{\nu_{\mu}}\right)d^{3}\mathbf{q}\Big]\,\,.
\label{antigenericboltz}
\end{eqnarray}
Neutrino and antineutrino evolution are nonlinearly coupled. The
collision terms differ for neutrinos and antineutrinos. Our new
generalized eight Boltzmann equations are entirely real-valued, yet
they contain all the quantum-mechanical oscillation phenomenology.

This is the complete set of kinetic equations for the real-valued
neutrino phase-space densities $f_{\nu_{e}}$ and $f_{\nu_{\mu}}$ and
the corresponding antineutrino phase-space densities
$\tilde{f}_{\nu_{e}}$ and $\tilde{f}_{\nu_{\mu}}$%
\footnote{For a slightly different way of writing the generalized
Boltzmann equations (in terms of the $B$'s in eq. (\ref{B's})), we
refer to the accompanying paper (Strack \& Burrows 2005).}.
In principle, one needs to solve the above set of eight real-valued
equations to simulate and understand neutrino evolution in
phase-space as for instance in numerical calculations for the
supernova explosion mechanism and the early universe. Our
generalized Boltzmann equations are intrinsically nonlinear due to
the neutrino self-interactions from $\Omega_{\nu\nu}$ and $\tilde{\Omega}_{\tilde{\nu}\nu}$.\\

Note that in the absence of collisions, Liouville's theorem for the
total neutrino flavor specific intensity,
\begin{eqnarray}
\frac{d}{dt}\left(f_{\nu_{e}}+f_{\nu_{\mu}}\right)=0\,\,,
\end{eqnarray}
where $\frac{d}{dt}$ denotes the total time derivative, is
recovered.\\

Moreover, at sufficiently high densities, one must include blocking
corrections for the collision terms as well as for the oscillation
terms which would further increase the nonlinearity of the
equations. We feel that this is straightforward; there is no need to
include these statistical properties in the formal framework.

For blocking corrections to be important for neutrinos, however, the
densities must be sufficiently high ($\rho>10^{8}$ g cm$^{-3}$). At
these densities observable neutrino oscillations are suppressed due
to the very small effective mixing angle in matter which can be
written as (Wolfenstein 1979, Pantaleone 1989):
\begin{eqnarray}
\sin2\theta=\frac{\sin2\theta_{vac}}{\sqrt{\sin^{2}2\theta_{vac}+(\cos2\theta_{vac}-\frac{A\varepsilon}{\Delta
m^{2}})^{2}}}\,\,. \label{mattermix}
\end{eqnarray}
In this expression for the effective mixing angle in matter,
$\theta_{vac}$ is the mixing angle for vacuum neutrino oscillations
and in numbers we have for the matter-induced effective mass term:
\begin{eqnarray}
\frac{A\varepsilon}{\Delta m^{2}}&=&
\frac{2\sqrt2G_{F}(Y_{e}/m_{n})\rho \varepsilon}{\Delta
m^{2}}\nonumber\\
& = & 1.53\times10^{-2}\left[\frac{Y_{e}\rho}{\text{g
cm}^{-3}}\right]\left[\frac{\varepsilon}{\text{MeV}}\right]\left[\frac{10^{-5}\text{eV}^{2}}{\Delta
m^{2}}\right]\,\,,
\end{eqnarray}
where $\rho$ denotes the mass density, $Y_{e}$ the number of
electrons per nucleon, and $m_{n}$ the nucleon mass.

One must include the other constituents of the ambient matter fields
such as neutrinos, nucleons and other leptons for an exact effective
mixing angle (Qian \& Fuller 1995). The density dependence is alike
and if anything, this additional contribution amplifies the effect
of matter suppression.
\subsection[Transport equations for the neutrino radiation
field]{Transport equations for the neutrino radiation field}
\label{transport}
It is instructive to rewrite eq. (\ref{genericboltz}) in terms of
the specific intensities of the neutrino radiation field. This way,
the collision integrals can be cast in a simple form. We neglect the
neutrino self-interactions in the following, having shown how to
include them in eq. (\ref{genericboltz}). Also we focus on the
neutrino degrees of freedom since without the self-interactions
neutrino and antineutrino evolution are coupled only in the normal
fashion through pair processes. We use the one-to-one relation
between the invariant phase-space densities, $f_{\nu_{e}}$,
$f_{\nu_{\mu}}$ and the specific intensities, $I_{\nu_{e}}$,
$I_{\nu_{\mu}}$ (Burrows, \etal\hspace{2mm}2000, Mihalas 1999) and
define
\begin{eqnarray}
I_{\nu_{e}}&=&\frac{\varepsilon^{3}f_{\nu_{e}}}{(2\pi
\hbar)^{3}c^{2}}\nonumber\\
I_{\nu_{\mu}}&=&\frac{\varepsilon^{3}f_{\nu_{\mu}}}{(2\pi
\hbar)^{3}c^{2}}\,\,.
\end{eqnarray}
The off-diagonal macroscopic overlap terms have the same units as
the phase-space densities and thus we may define their associated
specific intensities as
\begin{eqnarray}
\mathcal{R}_{e\mu}&=&\frac{\varepsilon^{3}f^{\mathcal{R}}_{e\mu}}{(2\pi
\hbar)^{3}c^{2}}\nonumber\\
\mathcal{I}_{e\mu}&=&\frac{\varepsilon^{3}f^{\mathcal{I}}_{e\mu}}{(2\pi
\hbar)^{3}c^{2}}\,\,. \label{reps}
\end{eqnarray}
These physically observable variables are commonly used in
hydrodynamic supernova simulations and we can write the extended set
of transport equations including neutrino oscillation phenomenology
in terms of the specific intensities.

The generalized Boltzmann equations in the laboratory (Eulerian
frame) for the radiation field of two oscillating neutrino species
with collisions are then:
\begin{eqnarray}
\frac{1}{c}\frac{\partial I_{\nu_{e}}}{\partial
t}+\frac{\mathbf{v}}{c}\cdot\frac{\partial
I_{\nu_{e}}}{\partial\mathbf{r}}
+\frac{\varepsilon^{3}\mathbf{\dot{p}}}{c}\cdot\frac{\partial\left(
I_{\nu_{e}}\varepsilon^{-3}\right)}{\partial \mathbf{p}}&=&
-\frac{2\pi}{L}\mathcal{I}_{e\mu}\sin2\theta+\mathcal{C}'_{\nu_{e}}\nonumber\\
\frac{1}{c}\frac{\partial I_{\nu_{\mu}}}{\partial
t}+\frac{\mathbf{v}}{c}\cdot\frac{\partial
I_{\nu_{\mu}}}{\partial\mathbf{r}}
+\frac{\varepsilon^{3}\mathbf{\dot{p}}}{c}\cdot\frac{\partial\left(
I_{\nu_{\mu}}\varepsilon^{-3}\right)}{\partial \mathbf{p}}&=&
\frac{2\pi}{L}\mathcal{I}_{e\mu}\sin2\theta+
\mathcal{C}'_{\nu_{\mu}}\nonumber\\
\frac{1}{c}\frac{\partial \mathcal{R}_{e\mu}}{\partial
t}+\frac{\mathbf{v}}{c}\cdot\frac{\partial
\mathcal{R}_{e\mu}}{\partial\mathbf{r}}+\frac{\varepsilon^{3}\mathbf{\dot{p}}}{c}\cdot\frac{\partial\left(
\mathcal{R}_{e\mu}\varepsilon^{-3}\right)}{\partial \mathbf{p}}&=&
-\frac{2\pi}{L}\left(\cos2\theta-A\right)\mathcal{I}_{e\mu}\nonumber\\
\frac{1}{c}\frac{\partial \mathcal{I}_{e\mu}}{\partial
t}+\frac{\mathbf{v}}{c}\cdot\frac{\partial
\mathcal{I}_{e\mu}}{\partial\mathbf{r}}
+\frac{\varepsilon^{3}\mathbf{\dot{p}}}{c}\cdot\frac{\partial\left(
\mathcal{I}_{e\mu}\varepsilon^{-3}\right)}{\partial
\mathbf{p}}&=&\frac{2\pi}{L}\left(
\frac{I_{\nu_{e}}-I_{\nu_{\mu}}}{2}\sin2\theta
+\left(\cos2\theta-A\right)\mathcal{R}_{e\mu}\right)\,\,.\nonumber\\
\label{radiationfield}
\end{eqnarray}
These equations and eqs. (\ref{genericboltz}) are our major results.
The collision terms can be conveniently written as (Burrows,
\etal\hspace{2mm}2000)
\begin{eqnarray}
\mathcal{C}'_{\nu_{e}}&=&-\kappa^{s}_{\nu_{e}}I_{\nu_{e}}+\kappa^{a
}_{\nu_{e}}\left(\frac{B_{\nu_{e}}-I_{\nu_{e}}}{1-\mathcal{F}^{eq}_{\nu_{e}}}\right)+
\frac{\kappa^{s}_{\nu_{e}}}{4\pi}
\int\Phi_{\nu_{e}}(\mathbf{\Omega},\mathbf{\Omega'})I_{\nu_{e}}(\mathbf{\Omega'})d\Omega'\,\,,
\end{eqnarray}
where $\Phi_{\nu_{e}}$ is a phase function for scattering into the
beam integrated over the solid angle $d\Omega'$. Furthermore,
$\kappa^{a }_{\nu_{e}}$ is the sum of all absorption processes
$\sum_{i}n_{i}\sigma^{a}_{i}$, where $n_{i}$ is the number density
of matter species $i$ and $\sigma^{a}_{i}$ denotes the absorption
cross sections (for scattering processes the superscript $a$ is
replaced with $s$).

$\mathcal{F}^{eq}_{\nu_{e}}$ is the equilibrium Fermi-Dirac
occupation probability
\begin{eqnarray}
\mathcal{F}^{eq}_{\nu_{e}}&=&\frac{1}
{\exp\left[\left(\varepsilon_{\nu_{e}}-\left(\mu_{e}-\hat{\mu}\right)\right)\beta\right]+1}\,\,,
\label{fermidirac}
\end{eqnarray}
where $\mu_{e}$ and $\hat{\mu}$ denote the electron chemical
potential and the difference between neutron and proton chemical
potential, respectively, and $B_{\nu_{e}}$ is the corresponding
black body specific intensity:
\begin{eqnarray}
B_{\nu_{e}}
&=&\frac{\varepsilon_{\nu_{e}}^{3}}{\left(2\pi\hbar\right)^{3}c^{2}}\mathcal{F}^{eq}_{\nu_{e}}.
\end{eqnarray}
Changing the subscript from $\nu_{e}$ to $\nu_{\mu}$ yields the
corresponding parameters for the $\nu_{\mu}$'s. For sterile
neutrinos, one substitutes $\nu_{s}$ for $\nu_{\mu}$ and sets the
scattering and absorption terms to zero. Similarly, one can write a
set of equations for antineutrinos with different collision terms
and with the sign of $A$ reversed. Here, neutrino and antineutrino
evolution are implicitly coupled through pair processes.
\section{Conclusions}
\begin{itemize}

\item We have formulated the most generic form of the generalized
Boltzmann equations including the nonlinear evolution in a neutrino
background in eq. (\ref{genericboltz}).\\

\item We have succeeded to incorporate the purely quantum-physical
phenomenon of flavor oscillations into the classical transport
equation (\ref{radiationfield}).\\

\item In the following, we employ the simple version of our new kinetic
equations in terms of the specific intensities, eq.
(\ref{radiationfield}), and perform explicit analytic and numerical
tests of this formalism in chapter \ref{decoherence}.\\

\item We show that our formalism is completely consistent with the
existing single-particle wavefunction approach
in subsection \ref{equivalent}.\\

\end{itemize}

\chapter[Simple Tests of the New Formalism]{Simple Tests of the New Formalism}
\label{decoherence}
In practice, it is interesting how the neutrino ensemble behaves
with time when there are flavor oscillations and collisions
simultaneously. Since the interaction rates for the differently
flavored neutrinos can differ dramatically--depending on the
surrounding matter constituents--, the possibility of flavor
oscillations must be taken into account for.

Eqs. (\ref{genericboltz}) and (\ref{radiationfield}) serve this
purpose. For the illustrative examples in the present chapter we
neglect the neutrino self-interactions, work with eq.
(\ref{radiationfield}), and demonstrate the correct limiting
behavior of the formalism.
\section[Flavor oscillations with absorptive matter
coupling]{Flavor oscillations with absorptive matter coupling}
As a first example, we solve eqs. (\ref{radiationfield}) for
$\nu_{e}$ -- $\nu_{\mu}$ oscillations in a box of isotropic
neutrinos that can also experience decohering absorption on matter.
This is realized by including the neutrino absorption reaction on
nucleons as given in Appendix \ref{appendixcross}. Every absorptive
collision decoheres the oscillation cycle. In reality, however, due
to the small cross-sections of neutrinos, frequent collisions with
the ambient matter fields require high densities. As indicated above
in eq. (\ref{mattermix}), observable flavor oscillations are
suppressed at the high densities found in stellar collapse
(Wolfenstein 1979). To remedy this problem artificially, we ``turn
off'' matter suppression by setting the matter interaction amplitude
$A$ of $\Omega_{mat}$ in eqs. (\ref{standardeffmass}) and
(\ref{mattermix}) to zero. This is not the situation found in
nature.

The goal of the following definitions is to simplify eq.
(\ref{radiationfield}) for the present purpose and write it in terms
of dimensionless variables. Therefore, we define the approximate
oscillation time,
\begin{eqnarray}
t_{osc}&\simeq&\frac{L}{2\pi c}=\frac{2\hbar\varepsilon}{\Delta
m^{2}c^{4}}\,\,,
\end{eqnarray}
and artificially set the interaction rate of the $\nu_{e}$'s equal
to the interaction rate of the $\nu_{\mu}$'s
\footnote{This definition of flavor-independent interaction rates is
again not so realistic. Cross sections depend on the mass of the
associated leptons which differ by a factor of $\approx200$.
Furthermore the abundance of the potential interaction partners in
the ambient matter is normally much greater for the $\nu_{e}$'s.}.
And we define the characteristic absorption time,
\begin{eqnarray}
t^{\nu_{e}}_{col}=\frac{1}{\kappa^{a
\ast}_{\nu_{e}}c}=\frac{\left(1-\mathcal{F}^{eq}_{\nu_{e}}\right)}{cN_{A}\rho
Y_{n}\sigma^a_{\nu_{e}n}}\,\,,
\end{eqnarray}
where $N_{A}$ denotes Avogadro's number, $Y_{n}$ is the neutron
fraction per nucleon, and $\sigma^a_{\nu_{e}n}$ is the cross section
for absorption on neutrons as given in Appendix \ref{appendixcross}.
The absorption time is flavor-independent in our toy calculation. As
a ``critical parameter" for the characteristic time evolution of the
system we define the ratio of the oscillation to the absorption
timescales:
\begin{eqnarray}
\alpha=\frac{t_{osc}}{t^{\nu_{e}}_{col}}\,\,,
\end{eqnarray}
and finally we define the dimensionless time coordinate
\begin{eqnarray}
\tau=\frac{t}{t_{osc}}\,\,.
\end{eqnarray}
Furthermore, we set $B_{\nu_{e}}=B_{\nu_{\mu}}=B_{\nu}$ and denote
the dimensionless specific intensities by
\begin{eqnarray}
\widehat{I}_{\nu_{e}}(\tau)&=&\frac{I_{\nu_{e}}}{B_{\nu}}\nonumber\\
\widehat{I}_{\nu_{\mu}}(\tau)&=&\frac{I_{\nu_{\mu}}}{B_{\nu}}\,\,,
\end{eqnarray}
and for the off-diagonal macroscopic overlap functions we write
\begin{eqnarray}
\widehat{\mathcal{R}}_{e\mu}(\tau)&=&\frac{\mathcal{R}_{e\mu}}{B_{\nu}}\nonumber\\
\widehat{\mathcal{I}}_{e\mu}(\tau)&=&\frac{\mathcal{I}_{e\mu}}{B_{\nu}}\,\,.
\end{eqnarray}
Note that the black-body function is flavor-independent, since we
take both neutrino flavors to be of the same energy (chemical
potentials have also been set to zero; they do not affect the
solutions significantly). The resulting dimensionless version of eq.
(\ref{radiationfield}) reads:
\begin{eqnarray}
\frac{\partial \widehat{I}_{\nu_{e}}}{\partial
\tau}&=&-\widehat{\mathcal{I}}_{e\mu}\sin2\theta+\alpha\left(1-\widehat{I}_{\nu_{e}}\right)\nonumber\\
\frac{\partial \widehat{I}_{\nu_{\mu}}}{\partial
\tau}&=&\widehat{\mathcal{I}}_{e\mu}\sin2\theta+\alpha\left(1-\widehat{I}_{\nu_{\mu}}\right)\nonumber\\
\frac{\partial \widehat{\mathcal{R}}_{e\mu}}{\partial
\tau}&=&-\widehat{\mathcal{I}}_{e\mu}\cos2\theta \nonumber\\
\frac{\partial \widehat{\mathcal{I}}_{e\mu}}{\partial
\tau}&=&\frac{\widehat{I}_{\nu_{e}}-\widehat{I}_{\nu_{\mu}}}{2}\sin2\theta+
\widehat{\mathcal{R}}_{e\mu}\cos2\theta\,\,.\nonumber\\
\label{dimlessneutrons}
\end{eqnarray}
Since here we consider isotropic neutrinos, it is easy to show that
the scattering in and out of the beam -- represented by the integral
over the phase-function -- and the scattering part of the extinction
coefficient cancel
\footnote{In a vacuum ($\alpha=0$) and with the initial conditions
$\widehat{I}_{\nu_{e}}=1$, $\widehat{I}_{\nu_{\mu}}=0$ (and
consequently no off-diagonal overlap terms at $\tau=0$), we obtain
\begin{eqnarray}
\widehat{I}_{\nu_{e}}(\tau)&=&1-\sin^{2}2\theta\sin^{2}\frac{\tau}{2}\nonumber\\
\widehat{I}_{\nu_{\mu}}(\tau)&=&\sin^{2}2\theta\sin^{2}\frac{\tau}{2}\nonumber\\
\widehat{\mathcal{R}}_{e\mu}(\tau)&=&\frac{1}{2}\sin2\theta\cos2\theta\left(\cos\tau-1\right)\nonumber\\
\widehat{\mathcal{I}}_{e\mu}(\tau)&=&\frac{1}{2}\sin2\theta\sin\tau\,\,.
\label{homo}
\end{eqnarray}
This behavior of the radiation field is unambiguously identical to
the probability density obtained by squaring the amplitude of a
single-neutrino wavefunction in a beam. The off-diagonal terms
representing the macroscopic overlap peak when mixing of $\nu_{e}$
and $\nu_{\mu}$ neutrinos is maximal and vanish when the ensemble is
single-flavored.}.\\

In matter and for the initial conditions
$\widehat{I}_{\nu_{\mu}}=\widehat{\mathcal{R}}_{e\mu}=\widehat{\mathcal{I}}_{e\mu}=0$
and $\widehat{I}_{\nu_{e}}\neq0$, one can derive an harmonic
oscillator equation for the early rate of evolution of
$\widehat{I}_{\nu_{e}}$:
\begin{eqnarray}
\frac{\partial^{2}\widehat{I}_{\nu_{e}}}{\partial
\tau^{2}}+\left(\frac{1}{2}-\alpha^{2}\right)\widehat{I}_{\nu_{e}}=\text{const}\,\,.
\end{eqnarray}
As expected, for $\alpha\ll1$, the early time dependence of the
solution is predominantly sinusoidal: neutrino oscillations
dominate. For $\alpha\gg1$, an exponential decay/increase dominates
with a timescale of $1/\alpha$. Further algebraic manipulations
convert eq. (\ref{dimlessneutrons}) into an inhomogeneous ordinary
differential equation of fourth order for the dimensionless
electron-neutrino specific intensity, valid for all $\tau$:
\begin{eqnarray} \frac{\partial^{4}\widehat{I}_{\nu_{e}}}{\partial
\tau^{4}}+2\alpha\frac{\partial^{3}\widehat{I}_{\nu_{e}}}{\partial
\tau^{3}}
+\left(\alpha^{2}+\cot^{2}2\theta+1\right)\frac{\partial^{2}\widehat{I}_{\nu_{e}}}{\partial
\tau^{2}}+\alpha\left(2\cot^{2}2\theta+1\right)\frac{\partial
\widehat{I}_{\nu_{e}}}{\partial
\tau}+\nonumber\\
\left(\alpha\cot2\theta\right)^{2}
\left(\widehat{I}_{\nu_{e}}-1\right)=0\,\,. \label{fourthorder}
\end{eqnarray}
\\

In the following, we evaluate eqs. (\ref{dimlessneutrons}) and
(\ref{fourthorder}) for three different regimes (without matter
suppression since the matter interaction amplitude for the
$\nu_{e}$, $A$, has been set to zero artificially). Initial
conditions are
$\widehat{I}_{\nu_{\mu}}=\widehat{\mathcal{R}}_{e\mu}=\widehat{\mathcal{I}}_{e\mu}=0$
and $\widehat{I}_{\nu_{e}}=0.8$ and from the large-mixing-angle
solution (LMA) (Fukuda, \etal\hspace{2mm}2002) we apply
$\sin2\theta=0.9$ and $\Delta m^{2}=6.9\times10^{-5}$ eV$^{2}$ .
\subsection{Oscillation-dominated regime: $\alpha\ll1\,\,$}
We tailor eq. (\ref{fourthorder}) according to the smallness of
$\alpha$ and write
\begin{eqnarray} \frac{\partial^{4}\widehat{I}_{\nu_{e}}}{\partial
\tau^{4}}+\left(\cot^{2}2\theta+1\right)\frac{\partial^{2}\widehat{I}_{\nu_{e}}}{\partial
\tau^{2}}=0\,\,,
\end{eqnarray}
where we have dropped all terms with $\alpha$. This is an
approximation -- even in this regime.

The time dependence of the solution to this equation is
predominantly sinusoidal with oscillation periods of the order of
unity.\\

\begin{figure}
\begin{center}
\includegraphics[width=110mm]{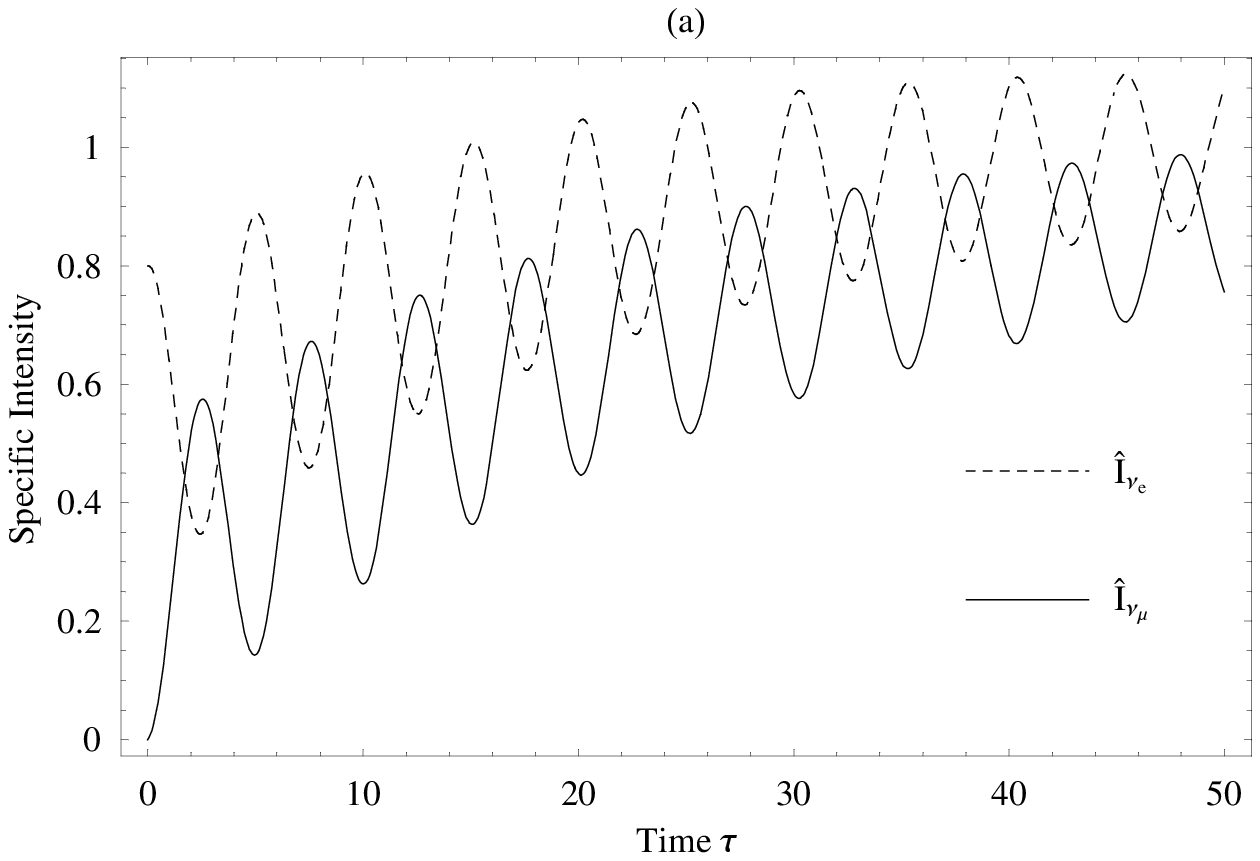}\\[5mm]
\includegraphics[width=110mm]{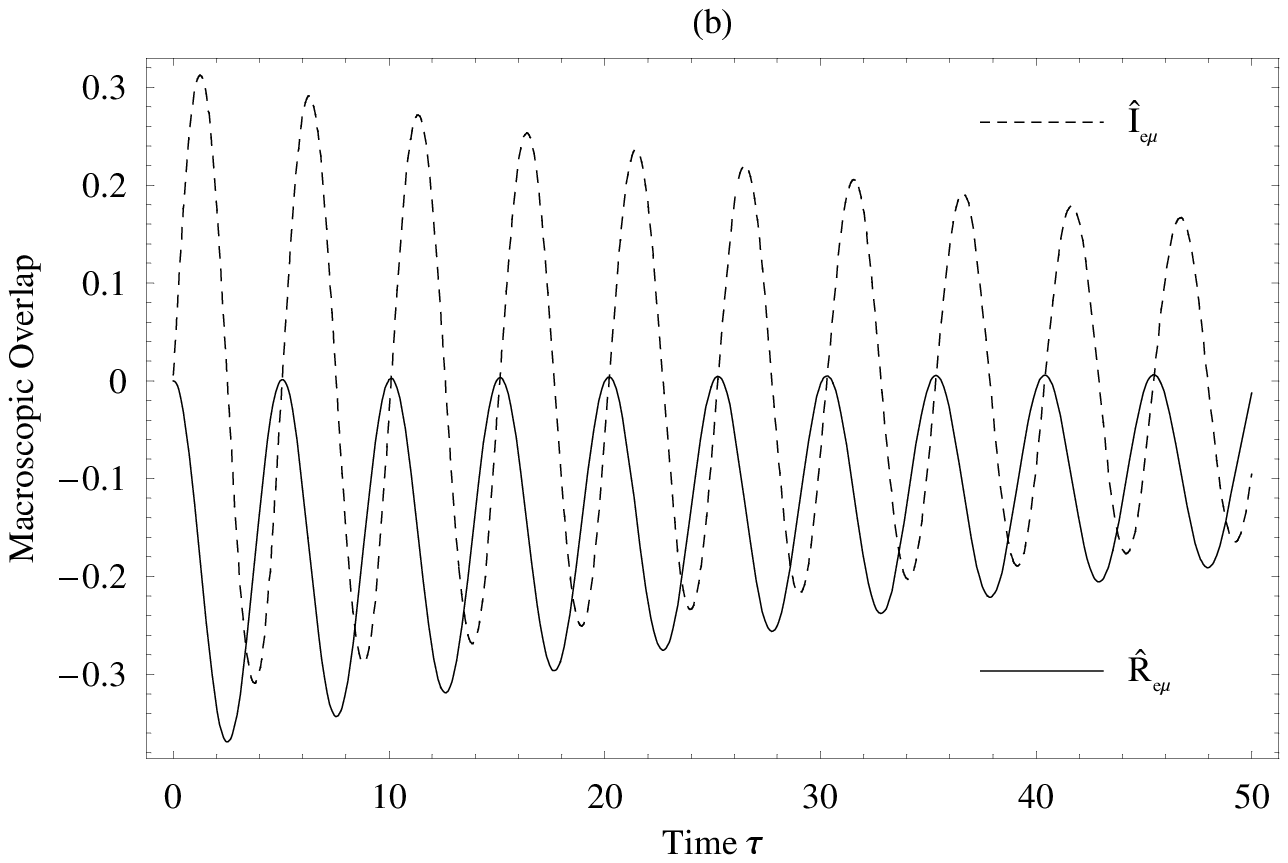}\\[5mm]
\caption{Oscillation-dominated regime. (\textbf{a}): dimensionless
specific intensities. (\textbf{b}): macroscopic overlap functions.
Parameters: $\varepsilon_{\nu_{e}}=\varepsilon_{\nu_{\mu}}=10$ MeV ,
$\rho=4\times 10^{11}\,\text{g cm}^{-3}$, and T $=1$ MeV.}
\label{fig:10hoch11}
\end{center}
\end{figure}

As depicted in Fig. \ref{fig:10hoch11}, flavor oscillations dominate
the temporal evolution of the neutrinos. Asymptotically, the
diagonal specific intensities for oscillating $\nu_{e}$ and
$\nu_{\mu}$ neutrinos equilibrate at 1.

Absorption on neutrons, and by detailed balance, the resulting
emissivity, drive the system to radiative equilibrium at the
blackbody intensity. The real part of the off-diagonal term,
$\widehat{\mathcal{R}}_{e\mu}$, takes negative values, whereas the
imaginary part, $\widehat{\mathcal{I}}_{e\mu}$, oscillates
symmetrically around zero. The oscillation amplitude of these terms
is comparable to the change in magnitude due to absorption. Both,
$\widehat{\mathcal{R}}_{e\mu}$ and $\widehat{\mathcal{I}}_{e\mu}$
converge toward zero as the ensemble becomes increasingly decohered;
the system eventually approaches flavor equilibrium and no
oscillations persist.

In other words, the density matrix of the system becomes diagonal
with time due to absorptive coupling with matter.
\newpage
\subsection{Oscillating and absorptive regime: $\alpha\simeq1\,\,$}
In this case, all terms of eq. (\ref{fourthorder}) have to be
retained. The solution to the homogenous equation consists of a sum
of exponentials with the four roots of the following equation in
Fourier space as exponents:
\begin{eqnarray}
\omega^{4}-2i\omega^{3}-\left(\cot^{2}2\theta+2\right)\omega^{2}+
i\left(1+2\cot^{2}2\theta\right)\omega+\cot^{2}2\theta=0\,\,,
\end{eqnarray}
where we have set $\alpha=1$. This equation has two roots of real
value and two roots of complex value with magnitudes of the order of
unity. Thus, decaying
exponentials and oscillating exponentials are represented equally.\\

\begin{figure}
\begin{center}
\includegraphics[width=110mm]{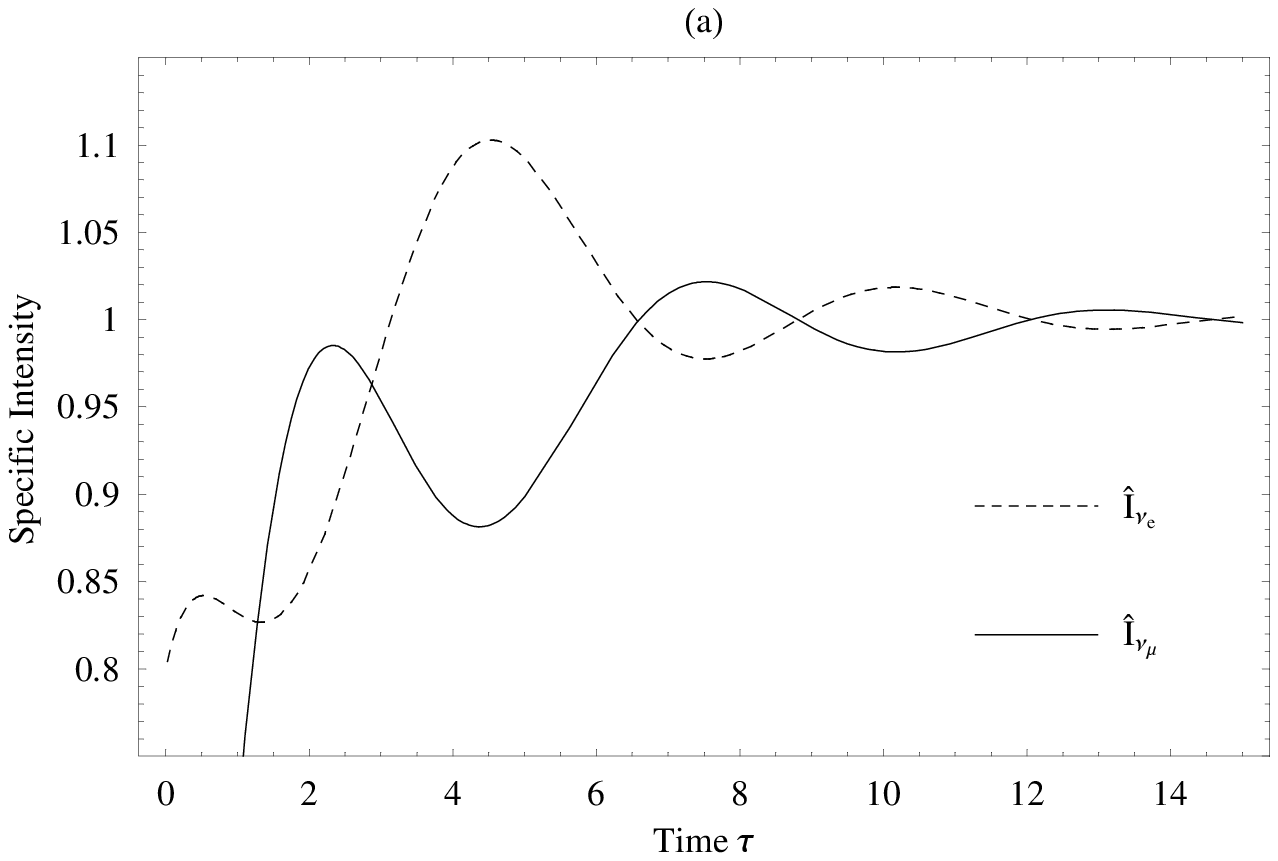}\\[5mm]
\includegraphics[width=110mm]{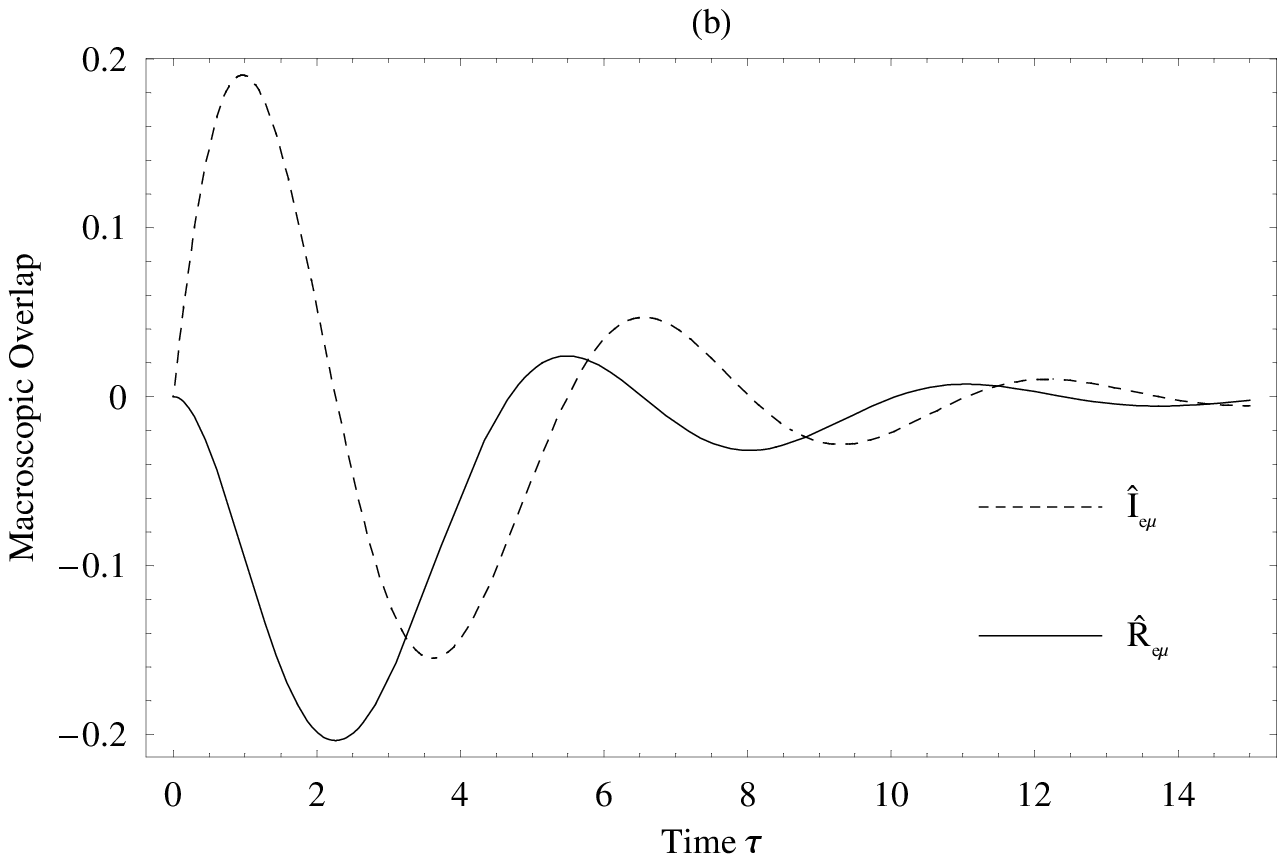}\\[5mm]
\caption{$\nu_{e}$ -- $\nu_{\mu}$ oscillations of isotropic
neutrinos in a box with absorptive matter coupling. (\textbf{a})
Specific intensities. (\textbf{b}) Off-diagonal macroscopic overlap
functions. Parameters:
$\varepsilon_{\nu_{e}}=\varepsilon_{\nu_{\mu}}=10$ MeV,
$\rho=8\times 10^{12}\,\text{g cm}^{-3}$, T $=5$ MeV.}
\label{fig:10hoch12}
\end{center}
\end{figure}

From Fig. \ref{fig:10hoch12}, it is clear that the equilibration due
to absorption happens on the same timescale as oscillations. The
ensemble is guided to flavor and radiative equilibrium. Coherent
flavor oscillations are disrupted by absorptive collisions.
Asymptotically, the diagonal specific intensities for the
$\nu_{e}$'s and $\nu_{\mu}$'s equilibrate.

Absorption on neutrons, and by detailed balance, the resulting
emissivity, drive the $\nu_{e}$'s to the blackbody intensity. The
oscillation amplitude decreases with time; the quantum evolution of
the system is decohered through absorptive coupling with matter.
This is sometimes referred to as quantum decoherence (Raffelt, Sigl
\& Stodolsky 1993). The real part of the off-diagonal overlap,
$\widehat{\mathcal{R}}_{e\mu}$, takes predominantly negative values
whereas the imaginary part, $\widehat{\mathcal{I}}_{e\mu}$,
oscillates symmetrically around zero. Both vanish asymptotically and
no oscillations persist.
\newpage
\subsection{Absorption-dominated regime: $\alpha\gg1\,\,$} We extract
the terms of eq. (\ref{fourthorder}) consisting of $\alpha$ and
write the approximate equation for this regime as
\begin{eqnarray}
\frac{\partial^{3}\widehat{I}_{\nu_{e}}}{\partial \tau^{3}}
+\frac{\alpha}{2}\frac{\partial^{2}\widehat{I}_{\nu_{e}}}{\partial
\tau^{2}}+ \frac{1}{2}\frac{\partial \widehat{I}_{\nu_{e}}}{\partial
\tau}+\frac{\alpha\cot^{2}2\theta}{2}
\widehat{I}_{\nu_{e}}-\frac{\alpha\cot^{2}2\theta}{2}=0\,\,.
\end{eqnarray}
The solution is predominantly exponentially waxing or waning
(depending on the initial conditions) with timescale $1/\alpha$.

\begin{figure}
\begin{center}
\includegraphics[width=110mm]{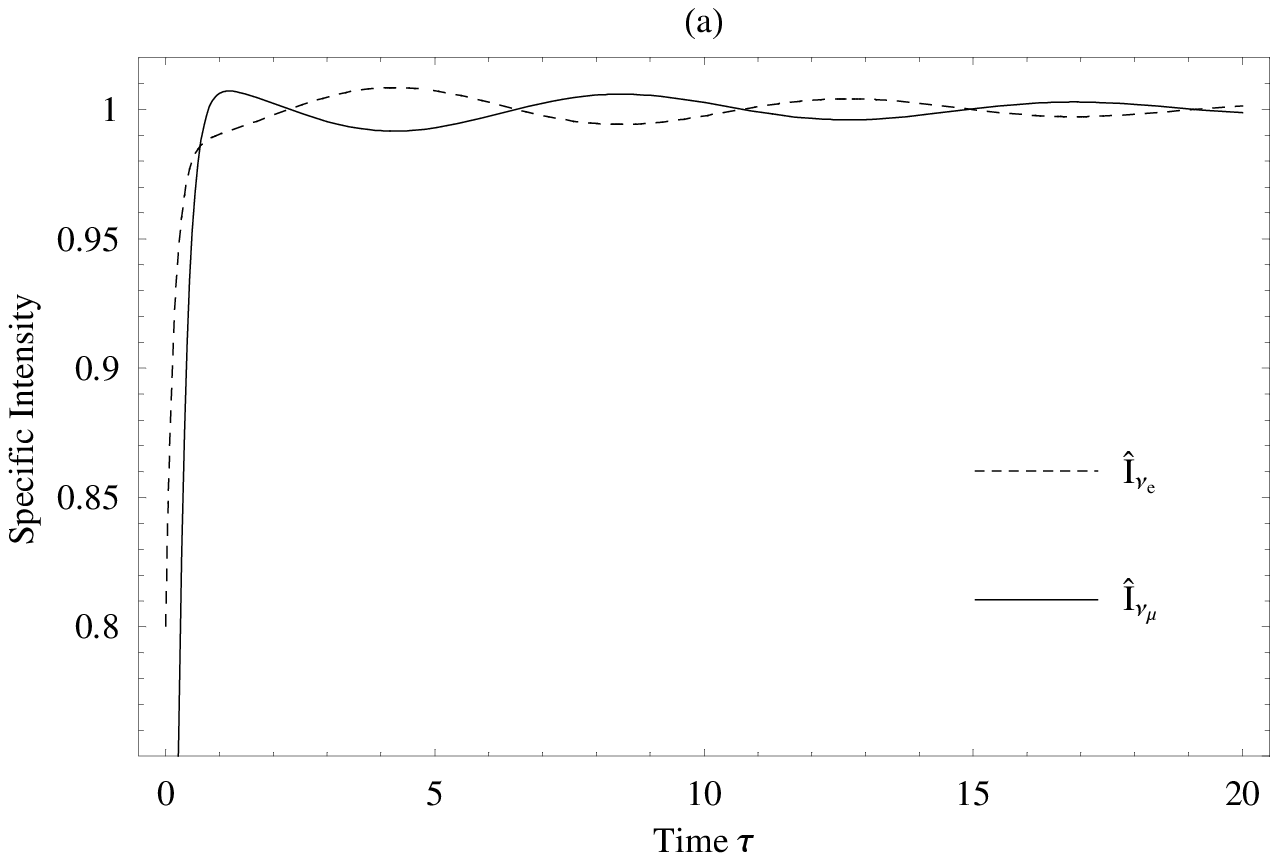}\\[5mm]
\includegraphics[width=110mm]{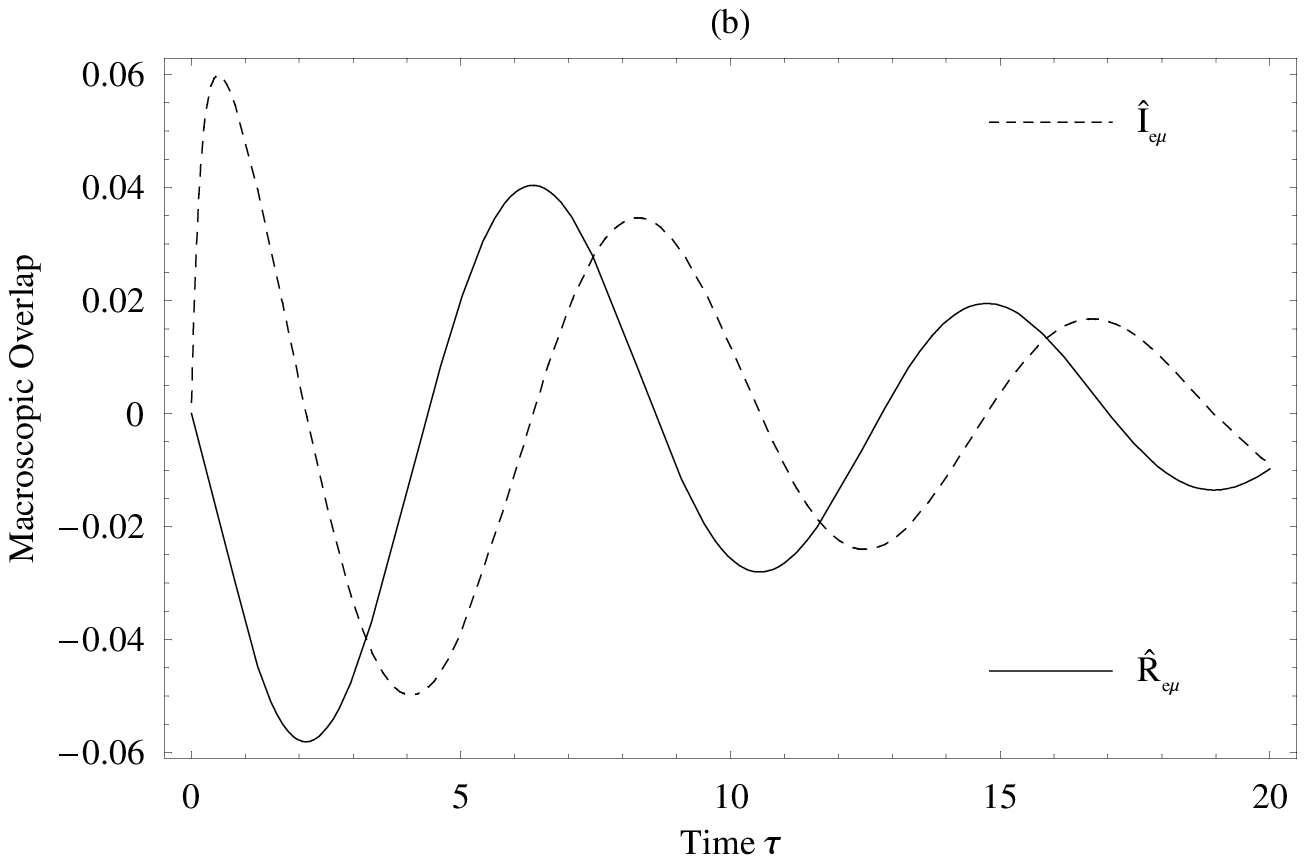}\\[5mm]
\caption{Absorption-dominated regime. (\textbf{a}): dimensionless
specific intensities. (\textbf{b}): off-diagonal macroscopic overlap
functions. Note the small values on the ordinate of (\textbf{b}).
Parameters: $\varepsilon_{\nu_{e}}=\varepsilon_{\nu_{\mu}}=10$ MeV ,
$\rho=4\times 10^{13}\,\text{g cm}^{-3}$, and T $=10$ MeV.}
\label{fig:10hoch13}
\end{center}
\end{figure}

The equilibration due to absorption is the dominant feature of this
regime as shown in Fig. \ref{fig:10hoch13}. Nonetheless, the
specific intensities execute small-amplitude oscillations around the
black-body value (a small effect due to neutrino mixing). The
diagonal specific intensities for $\nu_{e}$ and $\nu_{\mu}$
neutrinos equilibrate at the blackbody intensity in fractions of one
reduced oscillation time $\tau$. The amplitude of the corresponding
off-diagonal macroscopic overlap functions is very small (note the
values on the ordinate). In a nutshell, oscillations do not
influence the solution significantly.
\newpage
\section{Active-sterile oscillation in a beam with absorptive matter
coupling} \label{sec:sterile}
The mass of hypothetical sterile neutrinos can only be constrained
by observation (Abazajian, Fuller \& Patel 2001). It is commonly
thought that it could be much larger than the mass of the active
neutrino species resulting in a comparatively large
mass-squared-difference: $\Delta m^{2}\sim
\mathcal{O}(\text{eV}^{2})$ (Abazajian, Bell, Fuller \& Wong 2004).
This shifts the relevance of flavor oscillations to relatively high
densities.

As in the preceding section, we artificially ``turn off'' matter
suppression by setting the matter interaction amplitude $A$ to zero.
We determine the specific intensities of a $\nu_{e}$ neutrino beam
simultaneously oscillating into a sterile species $\nu_{s}$ and
being absorbed by nucleons including only the reaction in Appendix
\ref{appendixcross}.

To this end, we define the approximate ``reduced'' oscillation
length
\begin{eqnarray}
l_{osc}&\simeq&\frac{L}{2\pi
\sin2\theta}=\frac{2\hbar\varepsilon}{\Delta
m^{2}c^{3}\sin2\theta}\,\,,
\end{eqnarray}
and the characteristic absorption length
\begin{eqnarray}
l^{\nu_{e}}_{col}=\frac{1}{\kappa^{a
\ast}_{\nu_{e}}}=\frac{\left(1-\mathcal{F}^{eq}_{\nu_{e}}\right)}{N_{A}\rho
Y_{n}\sigma^a_{\nu_{e}n}}\,\,,
\end{eqnarray}
where $N_{A}$ denotes Avogadro's number, and $Y_{n}$ is the neutron
fraction per nucleon, and $\sigma^a_{\nu_{e}n}$ is the cross section
for absorption on neutrons. We define the ratio of oscillation to
absorption length scales
\begin{eqnarray}
\alpha=\frac{l_{osc}}{l^{\nu_{e}}_{col}}\,\,,
\end{eqnarray}
and the dimensionless distance coordinate
\begin{eqnarray}
\xi=\frac{x}{l_{osc}}\,\,.
\end{eqnarray}
We denote the dimensionless specific intensities and the
off-diagonal macroscopic overlap functions normalized to the
blackbody intensity with a hat.
\begin{figure}
\begin{center}
\includegraphics[width=110mm]{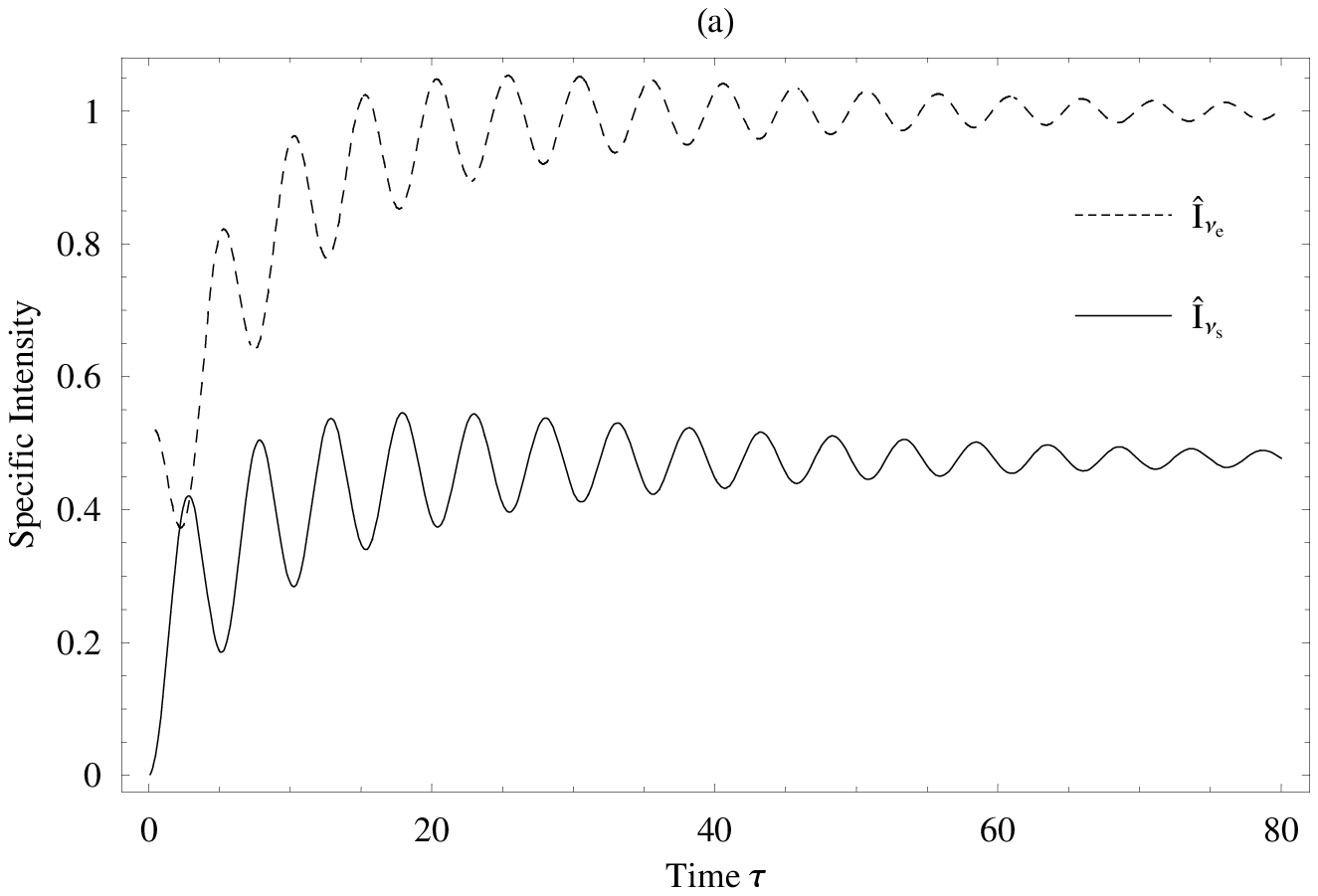}\\[5mm]
\includegraphics[width=110mm]{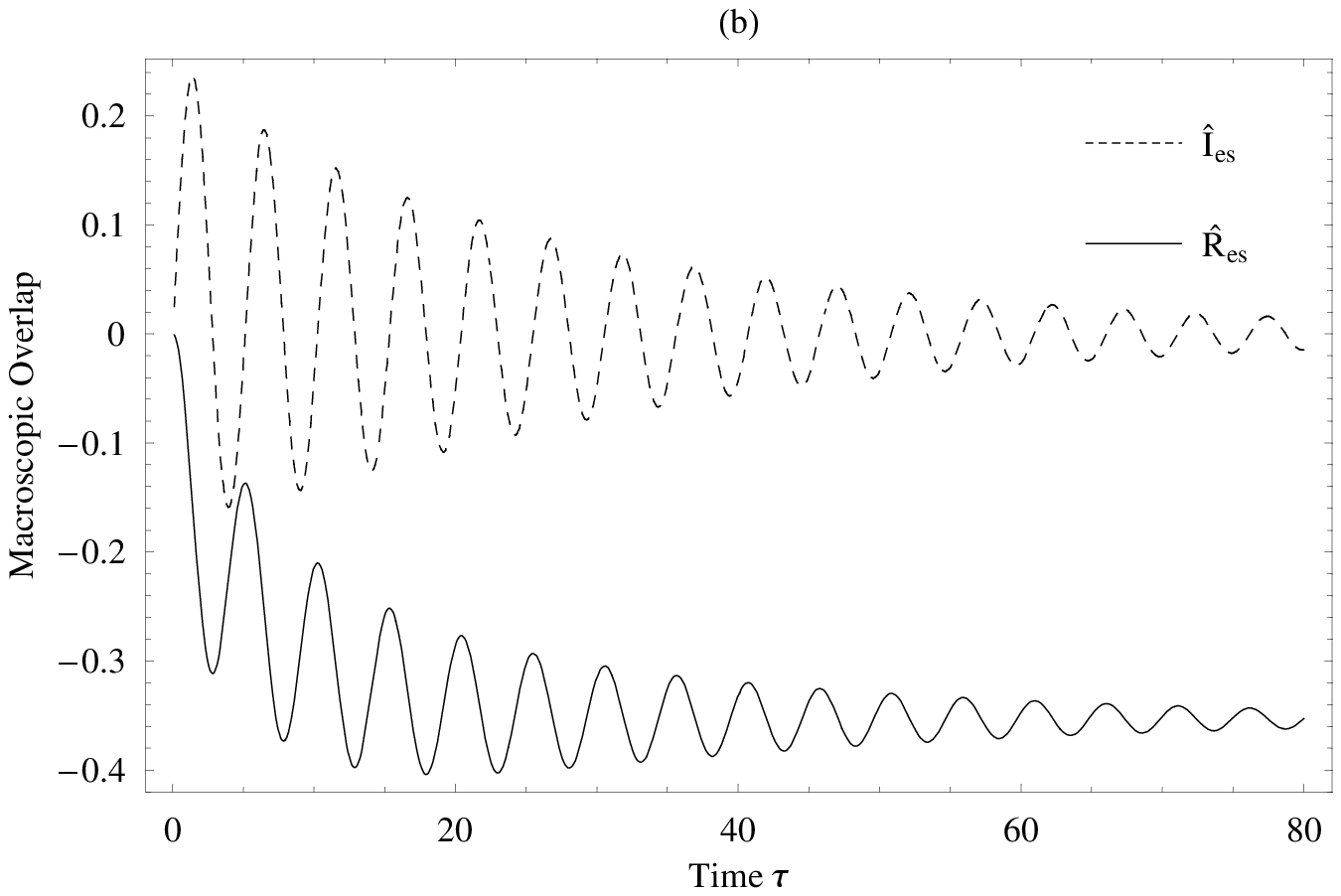}\\[5mm]
\caption{Active-sterile conversion with absorptive matter coupling.
(\textbf{a}) Specific intensities. (\textbf{b}) Off-diagonal
macroscopic overlap functions. Parameters:
$\varepsilon_{\nu_{e}}=\varepsilon_{\nu_{s}}=40$ MeV, $\rho=5\times
10^{13}\,\text{g cm}^{-3}$, $\sin2\theta=0.9$, T $=10$ MeV, and
$\Delta m^{2}=0.1$ eV$^{2}$ (Abazajian, Bell, Fuller \& Wong 2004).}
\label{fig:sterile}
\end{center}
\end{figure}
\newpage
The resulting dimensionless version of eq. (\ref{radiationfield})
reads:
\begin{eqnarray}
\frac{\partial \widehat{I}_{\nu_{e}}}{\partial
\xi}&=&-\widehat{\mathcal{I}}_{es}+\alpha\left(1-\widehat{I}_{\nu_{e}}\right)\nonumber\\
\frac{\partial \widehat{I}_{\nu_{s}}}{\partial
\xi}&=&\widehat{\mathcal{I}}_{es}\nonumber\\
\frac{\partial \widehat{\mathcal{R}}_{es}}{\partial
\xi}&=&-\widehat{\mathcal{I}}_{es}\cot2\theta \nonumber\\
\frac{\partial \widehat{\mathcal{I}}_{es}}{\partial
\xi}&=&\frac{\widehat{I}_{\nu_{e}}-\widehat{I}_{\nu_{s}}}{2}+
\widehat{\mathcal{R}}_{es}\cot2\theta\,\,.\nonumber\\
\label{dimlesssterilebeam}
\end{eqnarray}
Note that there are no collision terms for sterile neutrinos.\\

In Fig. \ref{fig:sterile}, we depict the solutions to eq.
(\ref{dimlessneutrons}) for a $\nu_{e}$ neutrino beam simultaneously
oscillating into a sterile species $\nu_{s}$ and being absorbed by
nucleons.

In this toy example, $\alpha=0.2$. Initial conditions are
$\widehat{I}_{\nu_{s}}=\widehat{\mathcal{R}}_{es}=\widehat{\mathcal{I}}_{es}=0$
and $\widehat{I}_{\nu_{e}}=0.5$ . Flavor oscillations and collisions
happen on the same length scale. While passing through matter the
beam is guided to flavor equilibrium by the radiative equilibration.
The coherence of flavor oscillations is decreased with time. Similar
to the neutrinos in a box, the $\nu_{e}$'s asymptotically
equilibrate at the blackbody intensity. The sterile neutrinos in the
beam converge toward the value $0.475$ .

The interacting $\nu_{e}$'s indirectly govern the oscillatory
pattern of the $\nu_{s}$'s even though the sterile neutrinos do not
interact. The oscillation amplitude decreases with time; the quantum
evolution of the system is damped through absorptive coupling with
matter. The real part of the off-diagonal overlap,
$\widehat{\mathcal{R}}_{es}$, converges toward the finite negative
value $-0.35$ . The imaginary part, $\widehat{\mathcal{I}}_{es}$,
oscillates symmetrically around zero and asymptotically vanishes.
\newpage
\section[Matter-enhanced resonant flavor
conversion]{Matter-enhanced resonant flavor conversion}
\label{section:msw}
When a mixed neutrino traverses matter at certain densities, its
flavor composition may be resonantly altered. This is due to the,
compared to the vacuum, large difference between the effective
neutrino masses in matter. The $\nu_{e}$ receives an additional
effective mass contribution due its (compared with the other
flavors) super-allowed interaction rates. This has been incorporated
into the equations in this thesis; see eq.
(\ref{standardeffmass}).\\

To demonstrate that our formalism contains the MSW effect
(Wolfenstein 1979, Mikheyev \& Smirnov 1986, Bethe 1986), we solve
eq. (\ref{radiationfield}) for a monoenergetic one-dimensional
neutrino beam propagating down a density profile for which resonant
matter-enhanced flavor conversion takes place.

We define the dimensionless distance coordinate in terms of the
oscillation length
\begin{eqnarray}
\widehat{x}=x\,\frac{2\pi}{L},
\end{eqnarray}
and the dimensionless matter-induced mass term in terms of its
resonance value:
\begin{eqnarray}
\widehat{A}=\frac{A}{A_{res}}=\frac{A}{\cos2\theta}\,\,.
\label{dimlessvariables2}
\end{eqnarray}
The beam passes the resonance density for $\widehat{A}=1$. We
analyze the following dimensionless version of eq.
(\ref{radiationfield}):
\begin{eqnarray}
\frac{\partial I_{\nu_{e}}}{\partial\widehat{x}}&=&-\mathcal{I}_{e\mu}\sin2\theta\nonumber\\
\frac{\partial I_{\nu_{\mu}}}{\partial \widehat{x}}&=&\mathcal{I}_{e\mu}\sin2\theta\nonumber\\
\frac{\partial\mathcal{R}_{e\mu
}}{\partial \widehat{x}}&=&-\cos2\theta\left(1-\widehat{A}\right)\mathcal{I}_{e\mu}\nonumber\\
\frac{\partial\mathcal{I}_{e\mu}}{\partial
\widehat{x}}&=&\frac{I_{\nu_{e}}-I_{\nu_{\mu}}}
{2}\sin2\theta+\cos2\theta\left(1-\widehat{A}\right)\mathcal{R}_{e\mu}\,\,,\nonumber\\
\label{eq:msw}
\end{eqnarray}
where $\mathcal{C}'_{\nu_{e}}$ and $\mathcal{C}'_{\nu_{\mu}}$ have
been set to zero.\\

\begin{figure}
\begin{center}
\includegraphics[width=110mm]{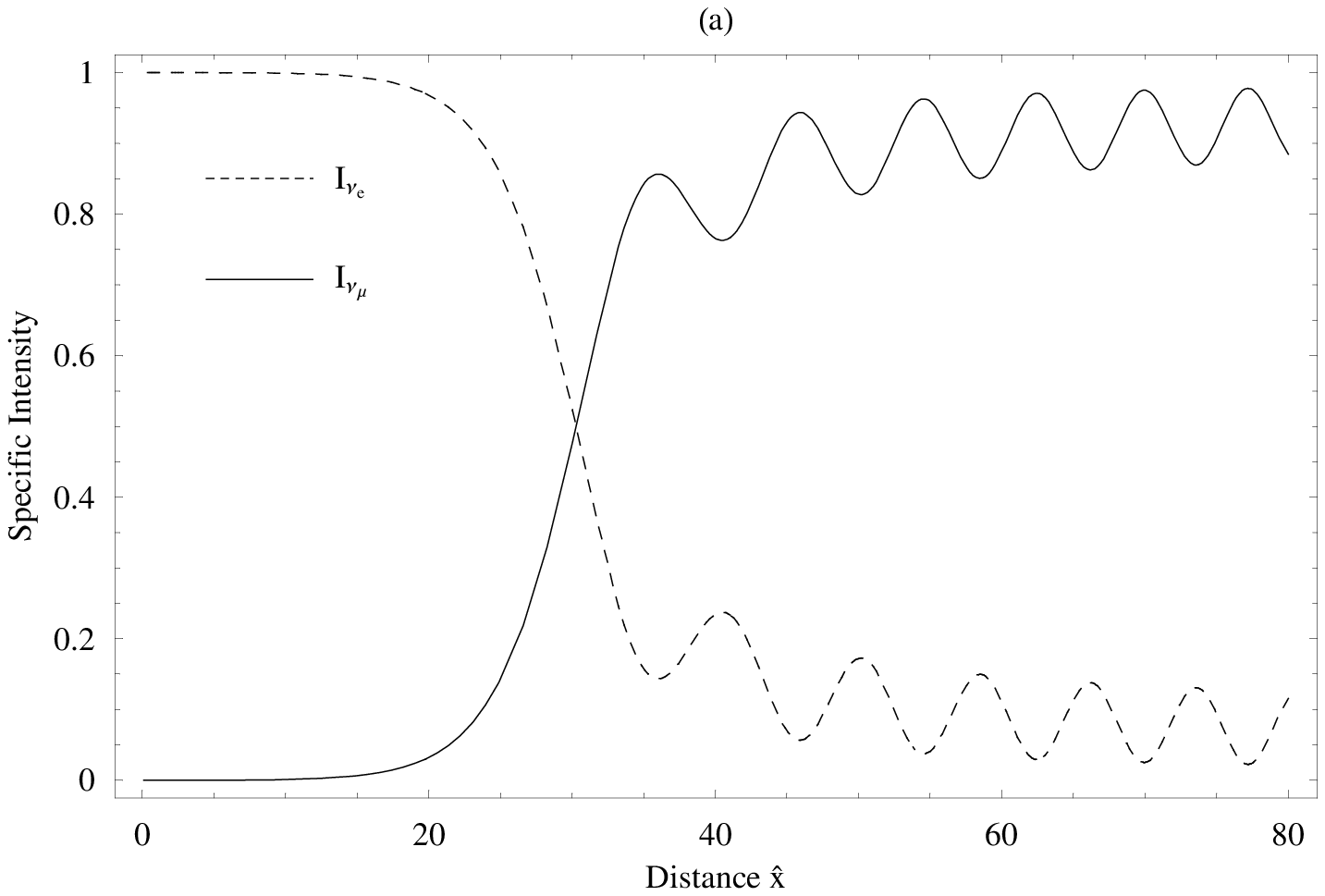}\\[5mm]
\includegraphics[width=110mm]{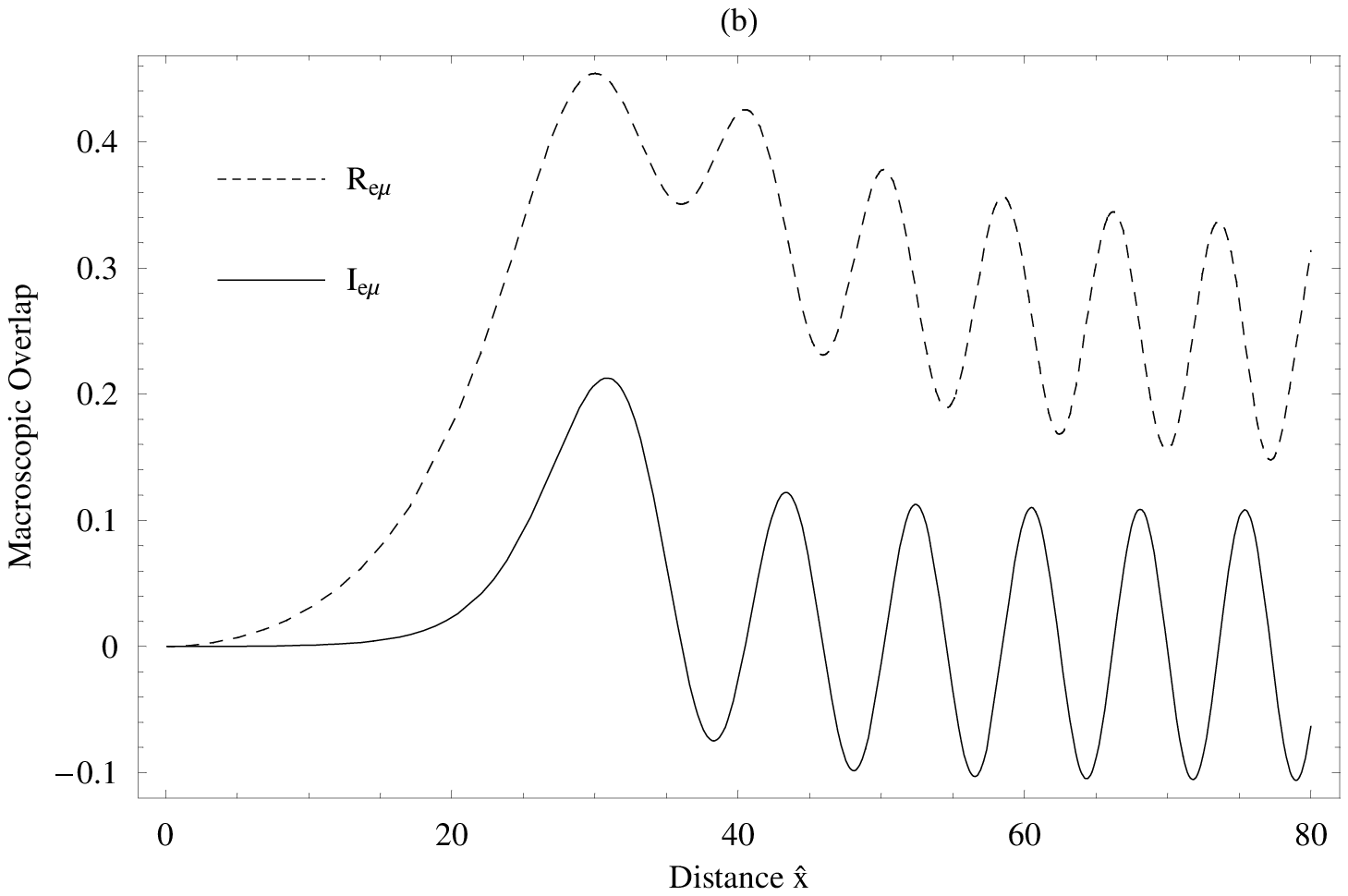}\\[5mm]
\caption{MSW effect. (\textbf{a}) Specific intensities. (\textbf{b})
Off-diagonal macroscopic overlap functions. An initially
$\nu_{e}$-beam propagates down the density profile
$\widehat{A}=\frac{\beta}{\widehat{x}^{2}}$, where $\beta=900$.}
\label{fig:msw}
\end{center}
\end{figure}

In Fig. \ref{fig:msw}, we depict the solutions to eq. (\ref{eq:msw})
for a density profile of
$\widehat{A}=\frac{\beta}{\widehat{x}^{2}}$. We set $\beta=900$ and
thus ensure that the scale of spatial inhomogeneities is large
compared to the microscopic length scales such as the neutrino de
Broglie wavelength and the oscillation length. Initially, the beam
contains only $\nu_{e}$ neutrinos. The mixing angle is arbitrarily
taken to be $\sin^{2}2\theta=0.18$ (at present the LMA is favored
(Fukuda, \etal\hspace{2mm}2002). From Fig. \ref{fig:msw}, it is
clear that at the resonance density, $\widehat{x}=\sqrt{\beta}=30$,
the flavor composition
of the beam is radically altered. For higher values of
$\widehat{x}$, the beam executes vacuum oscillations. In this
illustrative problem the density at production is much greater than
the resonance density. Then, the spatially averaged survival
probability of a $\nu_{e}$ neutrino going from matter to free space
should be (Parke 1986)
\begin{eqnarray}
\langle
P(\nu_{e}\rightarrow\nu_{e})\rangle\approx\left(1-P_{x}\right)\sin^{2}\theta+P_{x}\cos^{2}\theta\,\,,
\label{survival}
\end{eqnarray}
where $P_{x}$ is the Landau-Zener probability for non-adiabatic
transitions. For the chosen density profile, propagation is
adiabatic and $P_{x}=0$. The averaged $\nu_{e}$ survival probability
in Fig. \ref{fig:msw} converges toward $\sim 0.05$, which is
congruent with the value predicted in eq. (\ref{survival}):
$\sin^{2}\theta\sim0.05$. For high densities ($\widehat{x}\leq20$),
matter suppression is severe. No flavor oscillations happen and the
off-diagonal overlap functions $\mathcal{R}_{e\mu}$ and
$\mathcal{I}_{e\mu}$ are close to zero. In free space for
$\widehat{x}>60$, the imaginary part $\mathcal{I}_{e\mu}$ oscillates
symmetrically around zero whereas the real part $\mathcal{R}_{e\mu}$
is positive.
\subsection{Equivalence with wavefunction quantum mechanics}
\label{equivalent}
The solution given in Fig. \ref{fig:msw} is numerically equivalent
to the solution obtained using the standard wavefunction formalism
(Giunti 2004, Wolfenstein 1978, Bethe 1986, Mikheyev \& Smirnov
1986):
\begin{eqnarray}
i\frac{\partial}{\partial \widehat{x}}\left(\begin{matrix}
\psi_{\nu_{e}}\\
\psi_{\nu_{\mu}}\end{matrix}\right)&=&\frac{1}{2}\left(\begin{matrix}
-\cos2\theta\left(1-2\widehat{A}\right)&
\sin2\theta\\
\sin2\theta&\cos2\theta\end{matrix}\right) \left(\begin{matrix}
\psi_{\nu_{e}}\\
\psi_{\nu_{\mu}}\end{matrix}\right)\nonumber\\
-i\frac{\partial}{\partial \widehat{x}}\left(\begin{matrix}
\psi^{\ast}_{\nu_{e}}\\
\psi^{\ast}_{\nu_{\mu}}\end{matrix}\right)&=&\frac{1}{2}\left(\begin{matrix}
-\cos2\theta\left(1-2\widehat{A}\right)&
\sin2\theta\\
\sin2\theta&\cos2\theta\end{matrix}\right) \left(\begin{matrix}
\psi^{\ast}_{\nu_{e}}\\
\psi^{\ast}_{\nu_{\mu}}\end{matrix}\right)\,\,, \label{schroed}
\end{eqnarray}
when we identify the specific intensities with the probability
densities
\begin{eqnarray}
I_{\nu_{e}}&\leftrightarrow&|\psi_{\nu_{e}}|^{2}\nonumber\\
I_{\nu_{\mu}}&\leftrightarrow&|\psi_{\nu_{\mu}}|^{2}\,\,,
\label{match1}
\end{eqnarray}
and the macroscopic overlap functions with
\begin{eqnarray}
\mathcal{R}_{e\mu}&\leftrightarrow&\frac{1}{2}\left(\psi_{\nu_{e}}\psi^{\ast}_{\nu_{\mu}}
+\psi^{\ast}_{\nu_{e}}\psi_{\nu_{\mu}}\right)\nonumber\\
\mathcal{I}_{e\mu}&\leftrightarrow&\frac{1}{2i}\left(\psi_{\nu_{e}}\psi^{\ast}_{\nu_{\mu}}
-\psi^{\ast}_{\nu_{e}}\psi_{\nu_{\mu}}\right)\,\,. \label{match2}
\end{eqnarray}
Thus, our Boltzmann formalism is completely consistent with the
existing description.
\newpage
\section{Conclusions}
In this chapter, we have demonstrated that our generalized Boltzmann
equations for the neutrino specific intensities, eq.
(\ref{radiationfield}), reproduce the expected behavior. We have
considered three different settings, for which we numerically solved
dimensionless versions of eq. (\ref{radiationfield}).\\

\begin{itemize}

\item The vacuum beam solution was calculated in
analytic form for a non-interacting neutrino ensemble in a box, eq.
(\ref{homo}).\\

\item We have shown how the interplay between flavor oscillation
and decoherent matter coupling can be analyzed simply be solving eq.
(\ref{dimlessneutrons}) for neutrinos undergoing absorptions on
nucleons. The graphical results for three different regimes are
given in Figs. \ref{fig:10hoch11}, \ref{fig:10hoch12}, and
\ref{fig:10hoch13}.\\

\item An important feature of our new approach is that the MSW effect for a streaming
neutrino beam can be analyzed in a very simple fashion for
observable quantities, see eq. (\ref{eq:msw}). Numerical results for
an exemplary density profile are depicted in Fig. \ref{fig:msw}.\\

\item For realistic physical settings in a supernova environment, one
must take into account the energy spectrum of the neutrinos, energy
redistribution scattering, etc., and calculate the solutions to eq.
(\ref{radiationfield}) in their full generality. This has not been
performed in this work and it would also be interesting to
understand better the effects of the self-interaction contributions
on the right-hand-side of eq. (\ref{genericboltz}).\\

\end{itemize}
\chapter[Summary of Part I]{Summary of Part I}
We briefly highlight the main features of our formalism in flat
spacetime. The analytic expressions and conclusions reached in Part
1 are published in Physical Review D (Strack \& Burrows 2005).
\section{Quantum effects}

\begin{itemize}

\item Neutrinos are mathematically represented by spinor fields. One
consequence of their spin-$\frac{1}{2}$ nature is the Fermi-Dirac
state occupation probability. We include this in the stimulated
absorption terms of our formalism in eq. (\ref{fermidirac}).
Furthermore, we argue that it is straightforward to include blocking
corrections for the sink and source terms originating in the flavor
oscillations in eqs. (\ref{genericboltz}) and (\ref{radiationfield})
as well as in the corresponding collision integrals (Burrows,
\etal\hspace{2mm}2004).\\

\item In the simultaneous presence of collision and oscillations, two time scales
govern the evolution of the ensemble. When the collision or
interaction time is of the same order as the flavor oscillation
periods, it is believed that the collisions decohere the oscillation
cycles and reset their clock. Then the temporal evolution becomes
non-trivial since both processes must be taken into account for. We
have shown in chapter \ref{decoherence} and in particular the in
series of graphs, Figs. \ref{fig:10hoch11}, \ref{fig:10hoch12}, and
\ref{fig:10hoch13}, that our formalism encapsulates this interplay
of flavor oscillations and matter coupling in a straightforward way.\\

\item The possibility of oscillation into a sterile neutrino species
is explored in section \ref{sec:sterile}. We can observe indirect
affection of the sterile neutrinos, since their oscillating flavor
partner, the $\nu_{e}'s$ do interact. The plots are shown in Fig.
\ref{fig:sterile}. This is potentially important for kinetic
calculations in the early universe (Dolgov 1990) and could also have
an effect on the flavor composition of the neutrino spectra emerging
from supernovae (Abazajian,
\etal\hspace{2mm}2004).\\

\item Abstractly speaking, the neutrino vector is resonantly rotated in flavor space through
matter interaction with electrons (MSW effect) at certain densities.
This has been suggested to explain the solar neutrino puzzle (Bethe
1986). We have embedded this effect in a simple quasi-classical
framework and have shown analytically that our formalism correctly
displays this important quantum-physical prediction in subsection
\ref{equivalent}.
The corresponding plot is depicted in Fig. \ref{fig:msw}.\\

\end{itemize}

\section{Classical framework}

\begin{itemize}

\item The classical framework and the classical equations of radiation
hydrodynamics (Mihalas 1999) are commonly employed to simulate
core-collapse supernovae. The neutrinos are believed to play a
crucial role in the explosion process as they carry away almost all
the gravitational energy. Future detection of neutrino spectra from
supernova explosions will unravel the supernova mystery a little
more and buttress or falsify the theoretical models on the market.
Therefore it is very important to consistently include neutrino
oscillations into the propagation from the supernova to the detector
but also to include it in the kinetic description close to the
core.\\

\item Our generalized set of kinetic equations, eqs.
(\ref{genericboltz}) and (\ref{radiationfield}), will allow
straightforward extension of any numerical code that simulates
core-collapse, to account for the purely quantum-mechanical effects
of flavor oscillations and self-interactions. The generalized
Boltzmann equations without neutrino self-interactions, eq.
(\ref{radiationfield}), are very
simple and strongly resemble the classical equations of transport theory.\\

\item Conceptually, we have reduced the time evolution of neutrino
field operators over Fock space, eq. (\ref{wignerop}), to the time
evolution of real numbers by ensemble-averaging, eq.
(\ref{matrixelements}).\\

\item We have connected three approaches: We started with quantum
field operators, eq. (\ref{wignerop}), made the connection the
equations for the quasi-classical phase-space densities and the
associated specific intensities in section \ref{transport}, and
finally we showed that our approach is also consistent with the
one-particle wavefunction quantum mechanics
in subsection \ref{equivalent}.\\

\end{itemize}
\part[Curved Spacetime]{Curved Spacetime} \label{curved}
\chapter[Neutrino Phenomenology]{Neutrino Phenomenology}
\label{pheno}
In general astrophysical contexts such as the early universe and in
the vicinity of supernova cores, one needs to consider the effects
of gravitational fields on the physics under consideration. Here we
seek to understand the kinetics of oscillating neutrinos interacting
with a background medium in the presence of a strong, non-negligible
gravitational field.\\

Two changes to the eq. (\ref{genericboltz}) for flat spacetime need
be implemented. Firstly, we must analyze a gravitational
contribution to the mixing Hamiltonian (Cardall \& Fuller 1997),
which has observational consequences only if other effects such as
spin precession in the presence of a magnetic field take place
(Piriz, Roy \& Wudka 1996, Br\"uggen 1998). This gravitational
contribution stems from the spin connection, which we need to
consider when treating spinor fields in a curved spacetime (Brill \&
Wheeler 1957, Weinberg 1972).

In addition to the modification of the mixing Hamiltonian we must
write the left-hand-side of the Boltzmann equations, eq.
(\ref{genericboltz}), in a manifestly covariant manner so as to
account for curvature. To this end we follow (Mihalas 1999 \&
references therein, and Stewart 1971). This will be done in section
\ref{curvedboltz}.
\section{Neutrino oscillations $\left(\mathfrak{i}\right)$}
We discuss how mixed neutrinos can be analyzed in a covariant manner
following (Cardall \& Fuller 1997). We set $G=\hbar=c=1$ throughout
this section. In a standard treatment, the neutrino flavor state is
written (Kayser 1981)
\begin{eqnarray}
|\nu_{\alpha}(x,t)\rangle=\sum_{j}U_{\alpha j}
e^{-i\left(Et-P_{j}x\right)}|m_{j}\rangle\,\,,
\end{eqnarray}
where flavor (mass) indices are in Greek (Latin letters), and the
transformation coefficients are given by
\begin{eqnarray}
U=\left(
\begin{matrix}
     \sin \theta & \cos \theta \\
     -\cos \theta & \sin \theta
     \end{matrix} \right)\,\,.
\label{trafocoefficients}
\end{eqnarray}
The mass eigenstates are taken to be energy eigenstates with common
energy $E$. The three-momenta of the mass eigenstates are then
\begin{eqnarray}
P_{j}=\sqrt{E^{2}-m^{2}_{j}}\approx E-\frac{m^{2}_{j}}{2E}\,\,,
\label{momentumop}
\end{eqnarray}
where $m_{j}$ is the rest mass corresponding to the mass eigenstate
$|m_{j}\rangle$. To observe oscillation it is crucial that one
assumes null trajectory ($x=t$) of the mixed neutrino yielding the
expression
\begin{eqnarray}
|\nu_{\alpha}(x)\rangle=\sum_{j}U_{\alpha j}
\exp{\left(-i\frac{m_{j}}{2E}x\right)}|m_{j}\rangle\,\,.
\label{oscflat}
\end{eqnarray}
If the mass eigenstates could be measured at different positions (or
times), the interference pattern would be destroyed. How can eq.
(\ref{oscflat}) be generalized to curved spacetimes and be cast in a
manifestly covariant form?

The transformation coefficients from eq. (\ref{trafocoefficients})
(or equivalently the vacuum mixing angle $\theta$ between the two
flavor species of interest), the phase and the mass eigenstates are
frame independent quantities. The trajectory of a free neutrino is
null geodesic, which leads to
\begin{eqnarray}
|\nu_{\alpha}(\lambda)\rangle=\sum_{j}U_{\alpha j}
\exp{\left(i\int_{\lambda_{0}}^{\lambda}P^{\mu}p_{\mu}d\lambda\right)}|m_{j}\rangle\,\,.
\label{covariantphase}
\end{eqnarray}
In this expression, $P^{\mu}$ is the four-momentum operator that
generates spacetime translation of the mass eigenstates. The
quantity $p_{\mu}$ is the (null) tangent vector to the neutrino's
world line,
\begin{eqnarray}
x^{\mu}(\lambda)=\left(
\begin{matrix}
     x_{0}(\lambda)\\
     x_{1}(\lambda)\\
     x_{2}(\lambda)\\
     x_{3}(\lambda)
     \end{matrix} \right)\,\,,
\end{eqnarray}
and $\lambda$ is an affine parameter of the world line. In the
following, we prove that eq. (\ref{covariantphase}) is indeed the
proper expression for mixed neutrinos in a curved spacetime.

Therefore, we consider motion along the $x_{1}$ axis. Then the
four-momentum operator reads
\begin{eqnarray}
P^{\mu}=\left(E,P,0,0\right)\,\,, \label{1Dmomentumop}
\end{eqnarray}
and with the Lorentzian signature metric
\begin{eqnarray}
\eta_{\mu\nu}=\text{diag}\left[-1,1,1,1\right]\,\,,
\end{eqnarray}
we have for the argument of the exponential in eq.
(\ref{covariantphase})
\begin{eqnarray}
-i\int_{\lambda_{0}}^{\lambda}\left(E\frac{dt}{d\lambda}-P_{1}\frac{dx}{d\lambda}\right)d\lambda\,\,.
\end{eqnarray}
In this expression the $x_{1}$-component of the four-momentum
operator is
\begin{eqnarray}
P_{1}=\eta_{1\mu}P^{\mu}\cong E-\frac{M^{2}}{2E}\,\,.
\end{eqnarray}
$M$ is the mass operator which we replace by its eigenvalue and
write for the corresponding phase $\omega_{j}$ :
\begin{eqnarray}
\omega_{j}&=&-i\int_{\lambda_{0}}^{\lambda}\left[E\frac{dt/d\lambda}{dx/d\lambda}-
\left(E-\frac{m^{2}_{j}}{2E}\right)\right]
\frac{dx}{d\lambda}d\lambda\nonumber\\
&=&-i\int^{x}_{x_{0}}\frac{m^{2}_{j}}{2E}dx\nonumber\\
&=&-i\frac{m^{2}_{j}}{2E}\left(x-x_{0}\right)\,\,.
\end{eqnarray}
Here we have used the null trajectory condition
\begin{eqnarray}
\frac{dt/d\lambda}{dx/d\lambda}=\frac{p^{t}_{null}}{p^{1}_{null}}=1\,\,.
\end{eqnarray}
The important result is that eq. (\ref{covariantphase}) is the
covariant generalization of eq. (\ref{oscflat}) to spaces with
curvature.
\subsection{Matter-induced effective mass for the $\nu_{e}$}
\label{subsec:effmass}
We consider Dirac neutrinos and the equation of motion is the Dirac
equation. We can include matter effects into this equation. For this
purpose the Dirac equation in the flavor basis can be cast in the
form,
\begin{eqnarray}
\left[i\gamma^{\mu}\left(\partial_{\mu}+A_{\mu}\mathcal{P}_{L}\right)+\mathcal{M}\right]\psi=0\,\,.
\label{diracmatter}
\end{eqnarray}
Here $\psi$ is a column spinor with four components for each flavor.
$\gamma^{\mu}$ are defined by eq. (\ref{diracalgebra}) and our
choice shall be the chiral basis (Peskin \& Schroeder 1995):
\begin{eqnarray}
\gamma^{0}&=&\left(\begin{matrix}
0&1\\
1&0\end{matrix}\right)\,\,\nonumber\\
\gamma^{i}&=&\left(\begin{matrix}
0&\sigma^{i}\\
-\sigma^{i}&0 \end{matrix}\right)\,\,, \label{chiral}
\end{eqnarray}
where $\sigma^{i}$ are the Pauli-matrices. $\mathcal{P}_{L}$ is the
chirality projection on the left-handed components
\begin{eqnarray}
\mathcal{P}_{L}=\frac{1-\gamma^{5}}{2}\,\,,
\end{eqnarray}
where
\begin{eqnarray}
\gamma^{5}&=&-\frac{i}{4!}\epsilon^{\mu\nu\rho\sigma}\gamma_{\mu}\gamma_{\nu}\gamma_{\rho}\gamma_{\sigma}\nonumber\\
&=&\left(\begin{matrix}
-1&0\\
0 &1 \end{matrix}\right)\,\,.
\end{eqnarray}
Here one makes the assumption that only the left-handed components
of the flavor spinor $\psi$ interact, whereas the right-handed
components are taken to be sterile. The mass matrix in flavor basis
$\mathcal{M}$ satisfies,
\begin{eqnarray}
\mathcal{M}^{2}&=&U\left(\begin{matrix}
m^{2}_{1} & 0\\
0 & m^{2}_{2}
\end{matrix}
\right)U^{\dag}\,\,,
\end{eqnarray}
where $U$ has been specified in eq. (\ref{trafocoefficients}) and
involves the vacuum mixing angle $\theta$.

$A^{\mu}$ is the flavor-basis effective potential matrix for an
interaction with the electron background:
\begin{eqnarray}
A^{\mu}=-\sqrt{2}G_{F}\left(\begin{matrix}
N^{\mu}_{e} & 0\\
0 & 0
\end{matrix}
\right)\,\,. \label{effmass}
\end{eqnarray}
In this expression $N^{\mu}_{e}$ is the number current of the
electron fluid
\begin{eqnarray}
N^{\mu}_{e}=n_{e}u^{\mu}\,\,,
\end{eqnarray}
where $n_{e}$ is the electron density in the fluid rest frame, and
$u^{\mu}$ is the fluid's four-velocity. This is the covariant
extension of matrix given in eq. (\ref{mattermix}).

By iteration of the Dirac equation eq. (\ref{diracmatter}) in
Fourier space one has for the mass shell relation the expression,
\begin{eqnarray}
\left(P^{\mu}+A^{\mu}\mathcal{P}_{L}\right)\left(P_{\mu}+A_{\mu}\mathcal{P}_{L}\right)=-\mathcal{M}\,\,.
\end{eqnarray}
Keeping terms first order in $G_{F}$ and employing eq.
(\ref{1Dmomentumop}) we find
\begin{eqnarray}
P_{1}\simeq
E-\frac{1}{2E}\left(\mathcal{M}^{2}-\mathcal{V}\right)\,\,,
\label{3momentumeffective}
\end{eqnarray}
with
\begin{eqnarray}
\mathcal{V}=-2\sqrt{2}G_{F}\left(\begin{matrix}
E n_{e}\mathcal{P}_{L} & 0\\
0 & 0
\end{matrix}
\right)\,\,.
\end{eqnarray}
The three-momentum operator eq. (\ref{3momentumeffective}) now
includes effective mass contributions from background matter, and
can be used in eq. (\ref{covariantphase}).

The most important physical consequence of this effective mass is
well known (MSW effect) and we have performed explicit calculations
in this thesis in section \ref{section:msw}.
\subsection{Spinor fields}
Spinor fields and their transformation properties are investigated
with help of the vielbein formalism which separates local Lorentz
transformations and general coordinate transformations. This is
convenient because for spinor fields one can then apply, locally,
the spinor representation of the Lorentz group for which the well
known $\gamma$'s of the Dirac algebra in flat spacetime eq.
(\ref{diracalgebra}) can be employed. In Appendix \ref{vielbein}, we
sketch how to construct locally inertial coordinate systems and
``hide'' the curvature in the components of the vielbeins so as to
apply Lorentz transformations on spinors in curved spacetime
(Carroll 2004, Weinberg 1972, and Brill \& Wheeler 1957).

The Dirac equation in curved spacetime involves gravitational
effects on the spin through the spin connection $\Gamma_{\mu}$:
\begin{eqnarray}
\left(i\gamma^{a}e_{a}{}^{\mu}\left[\partial_{\mu}+\Gamma_{\mu}\right]+\mathcal{M}\right)\psi=0\,\,,
\label{diraccurved}
\end{eqnarray}
where Greek indices refer to the general coordinates, while the
Latin indices a, b, c, d refer to the locally inertial noncoordinate
basis as described in Appendix \ref{vielbein}. Using the explicit
expression for $\Gamma_{\mu}$ (Weinberg 1972):
\begin{eqnarray}
\Gamma_{\mu}=\frac{i}{8}\left[\gamma^{b},\gamma^{c}\right]e_{b}{}^{\nu}e_{c\nu;\mu}\,\,,
\label{gravvielbein}
\end{eqnarray}
where the semi-colon denotes the covariant derivative, one can
incorporate the gravitational effects on spin into the momentum
operator $P^{\mu}$ of eq. (\ref{covariantphase}). The Dirac equation
(\ref{diraccurved}) features the Dirac matrix product
\begin{eqnarray}
\gamma^{a}\left[\gamma^{b},\gamma^{c}\right]=2\eta^{ab}\gamma^{c}-2\eta^{ac}\gamma^{b}-
2i\epsilon^{dabc}\gamma_{5}\gamma_{d}\,\,,
\end{eqnarray}
where the right-hand-side can be calculated by using the identities
for Dirac matrices as for instance given in (Peskin \& Schroeder
1995). Here $\epsilon^{dabc}$ is the totally antisymmetric tensor
with $\epsilon^{0123}$=1. After some manipulation we can group the
terms arising from the spin connection as follows:
\begin{eqnarray}
i\gamma^{a}e_{a}{}^{\mu}\Gamma_{\mu}=i\gamma^{a}e_{a}{}^{\mu}\left(\mathfrak{G}_{\mu}
\left(-\frac{\gamma_{5}}{2}\right)\right)\,\,,
\end{eqnarray}
with
\begin{eqnarray}
\mathfrak{G}_{\mu}=\frac{i}{2}\epsilon^{dabc}\left(\gamma^{a}\right)^{-1}
\gamma_{d}e_{b}{}^{\nu}e_{c\nu;\mu}\,\,. \label{spineff}
\end{eqnarray}
Note that in the reference (Cardall \& Fuller 1997), a different
convention of the $\gamma$'s is employed and the corresponding
expression for $\mathfrak{G}_{\mu}$ is different (I could not see
how the authors derived their expression). It is now apparent that,
by adding a term proportional to the identity matrix (this has no
observable consequence), we can choose either chirality to be prone
to the gravitational effects and write
\begin{eqnarray}
i\gamma^{a}e_{a}{}^{\mu}\Gamma_{\mu}=i\gamma^{a}e_{a}{}^{\mu}\left(\mathfrak{G}_{\mu}
\mathcal{P}_{L}\right)\,\,,
\end{eqnarray}
where
\begin{eqnarray}
\mathcal{P}_{L}=\frac{1-\gamma^{5}}{2}
\end{eqnarray}
is the projector on the left-handed states.\\

An important point is that the gravitational contribution
$\mathfrak{G}^{\mu}$ is proportional to the identity matrix in
flavor space, and diagonal in spin space. It cannot induce spin
flips on its own. Therefore it will not have any observable effects
unless there are other off-diagonal terms in spin space, e.g.: from
the interaction of a neutrino magnetic moment with a magnetic field
(Piriz, Roy \& Wudka 1996). In reality, however, the spin-gravity
coupling needs a large anomalous magnetic moment of the
neutrino to yield observationally relevant effects.\\

At present the Standard Model-like calculation for massive Dirac
neutrinos predicts (Shrock \& Fujikawa 1977, Mohapatra \& Pal 2003)
\begin{eqnarray}
\mu\simeq\frac{3eG_{F}}{8\sqrt{2}\pi^{2}}m_{\nu}\,\,,
\end{eqnarray}
where $m_{\nu}$ is the mass of the neutrino. In numbers, we have
\begin{eqnarray}
\mu\simeq3.1\times10^{-19}\mu_{B}\left(\frac{m_{\nu}}{1
\text{eV}}\right)\,\,, \label{magmom}
\end{eqnarray}
$\mu_{B}$ being the Bohr magneton, $e/\left(2m_{e}\right)$. This is
rather puny. Even though the magnetic fields near protoneutron stars
or supernova cores can be as large as $10^{12}$ Gauss, the spin-flip
length scale will be nonlocal, i. e. spread out over large spatial
distances when compared with the scales of other processes as
collision with the ambient matter. For theoretical reasons it is
important that the gravitational contribution is added to the mixing
Hamiltonian eq. (\ref{mixhamilton}) even though with the present
range of magnetic fields and anomalous magnetic moments this will
have no physical effect.

To find effects (Lambiase 2004, Lambiase, Papini, Punzi \& Scarpetta
2005), the neutrino magnetic moment must be estimated to be
magnitudes larger than that following from eq. (\ref{magmom}) with
neutrino masses in the eV range.

\newpage
\section{Neutrino oscillations $\left(\mathfrak{ii}\right)$}
We can now treat the effective mass contribution of eq.
(\ref{effmass}) and the spin connection of eq. (\ref{spineff}) in
the same fashion. Then the flavor state in curved space time of eq.
(\ref{covariantphase}) can be written as
\begin{eqnarray}
|\nu_{\alpha}(\lambda)\rangle=\sum_{j}U_{\alpha j}
\exp{\left(i\int_{\lambda_{0}}^{\lambda}P^{\mu}p_{\mu}d\lambda\right)}|m_{j}\rangle\,\,.
\label{covariantphase2}
\end{eqnarray}
In this expression the argument of the exponential now encompasses
spin and effective mass effects and we have
\begin{eqnarray}
P^{\mu}p_{\mu}=-\left(\frac{\mathcal{M}^{2}}{2}+p^{\mu}\left(A_{\mu}+\mathfrak{G}_{\mu}\right)\mathcal{P}_{L}\right)\,\,.
\label{covariantphaseexplicit}
\end{eqnarray}
We are in position to explicitly calculate the vielbeins eq.
(\ref{gravvielbein}) for interesting geometries and follow the
temporal evolution of the flavor composition of the mixed neutrino
spinor in curved spacetime through solving the following modified
version of eq. (\ref{diraccurved}):
\begin{eqnarray}
\left(i\gamma^{a}e_{a}{}^{\mu}\left[\partial_{\mu}+\left(A_{\mu}+\mathfrak{G}_{\mu}\right)\mathcal{P}_{L}\right]
+\mathcal{M}\right)\psi=0\,\,. \label{diraccurvedmod}
\end{eqnarray}
More equations of this type are derived in the references Cardall \&
Fuller 1997, and Piriz, Roy \& Wudka 1996.
\section{Conclusions}
\begin{itemize}

\item In this chapter, we have discussed how to generalize the phase of a mixed neutrino to
curved spaces. The result is eq.
(\ref{covariantphaseexplicit}).\vspace{3mm}

\item We have shown the dynamical equation (\ref{diraccurvedmod}) for a neutrino spinor in curved spacetime.
We argued that the non-vanishing contribution due to spin-gravity
coupling can be written in terms of the vielbein fields, eq.
(\ref{spineff}). This contribution is diagonal in spin and flavor
space.\vspace{3mm}

\item The terms in the operator of eq.
(\ref{diraccurvedmod}), notably the gravitational contribution,
$\mathfrak{G}_{\mu}$, will be used to write the mixing Hamiltonian,
eq. (\ref{mixhamiltoncurved}), so as to account for spin-gravity
coupling.\vspace{3mm}

\end{itemize}

\chapter[General-relativistic Boltzmann Formalism]
{General-relativistic Boltzmann Formalism} \label{curvedformalism}

The purpose of section \ref{curvedboltz} is to formulate the
classical Boltzmann equation
\begin{eqnarray}
\frac{\partial f(\mathbf{r},\mathbf{p},t)}{\partial t}+
\dot{\mathbf{r}}\cdot\frac{\partial
f(\mathbf{r},\mathbf{p},t)}{\partial
\mathbf{r}}+\dot{\mathbf{p}}\cdot\frac{\partial
f(\mathbf{r},\mathbf{p},t)}{\partial \mathbf{p}}= \left[\frac{\delta
f}{\delta t}\right]_{col}\,\,, \label{classboltz}
\end{eqnarray}
in a manifestly covariant manner and generalize the Liouville
operator of the left-hand-side to curved spaces. In section
\ref{sec:curvedmixing}, we decompose the phase-space densities into
their flavor degrees of freedom and into their spin degrees of
freedom. We show the associated mixing Hamiltonian, eq.
(\ref{mixhamiltoncurved}).

The highlight of this chapter are the new kinetic equations for
oscillating neutrinos in curved spaces, eq. (\ref{eq:curvedboltz}).
\vspace{3mm}
\section{Classical Boltzmann equation}
\label{curvedboltz}
The phase-space for a system of relativistic particles is simply the
tangent bundle over the spacetime manifold, and the corresponding
distribution function or phase-space density can naturally be
described as a scalar field on this bundle (Stewart 1971). At every
point $x$ of the semi-riemannian manifold of our spacetime, the
tangent space, $T_{x}$, becomes the relativistic generalization of
momentum space. Physically realizable four-momenta are timelike or
null. Free particles have a geodesic trajectory and between
collisions, the four-momenta of the particles,
\begin{eqnarray}
\frac{dx^{\mu}}{d\lambda}=p^{\mu}\,\,,
\end{eqnarray}
where $\lambda$ is an affine parameter, satisfy the geodesic
equation:
\begin{eqnarray}
\frac{dp^{\mu}}{d\lambda}+\Gamma^{\mu}{}_{\nu\rho}p^{\nu}p^{\rho}=0\,\,.
\label{geodesic}
\end{eqnarray}
In this expression $\Gamma^{\mu}{}_{\nu\rho}$ is the symmetric,
metric-compatible and unique Christoffel connection that is linear
in the derivatives of the metric:
\begin{eqnarray}
\Gamma^{\mu}{}_{\nu\rho}=\frac{1}{2}g^{\mu\alpha}\left(\frac{\partial
g_{\alpha\nu}}{\partial x^{\rho}}+\frac{\partial
g_{\rho\alpha}}{\partial x^{\nu}}-\frac{\partial
g_{\nu\rho}}{\partial x^{\alpha}}\right) \label{christoffel}\,\,,
\end{eqnarray}
where $g_{\mu\nu}$ is the twice-covariant metric tensor field which
multi-linearly makes two tangent vectors to real numbers.\\

The manifestly covariant relativistic Boltzmann equation for the
invariant phase-space density $f\left(x^{\mu},p^{\mu}\right)$ is
then given by:
\begin{eqnarray}
p^{\mu}\frac{\partial{f}}{\partial
x^{\mu}}-\left(\Gamma^{\mu}{}_{\sigma\rho}p^{\sigma}p^{\rho}\right)\frac{\partial
f}{\partial p^{\mu}}=\left[\frac{\delta f}{\delta
\lambda}\right]_{coll}\,\,. \label{covboltz}
\end{eqnarray}
The term on the right-hand-side is the collision term (Mihalas
1999).

There is a subtlety involved in this way of writing the Boltzmann
equation. We have tacitly assumed that the invariant phase-space
density $f\left(x^{\mu},p^{\mu}\right)$ is defined for all possible
four-momenta. However only the four-momenta that lie on the
null-cone make sense for the description of almost massless
neutrinos (as in chapter \ref{pheno} we assume null geodesics for
the neutrinos). Therefore, we must ensure that $p^{\mu}$ remains on
the null cone as the neutrino propagates. This can be achieved by
employing the vielbein formalism (Appendix \ref{vielbein}) because
in the local vielbein frame we can fix the four-momentum to be null
once and for all, since in the vielbein frame we have the Minkowski
metric. All the spatial-dependence involving the curvature is
``out-sourced'' to the complicated structure of the vielbeins. We
regard the phase-space density $f$ as a function of $(x^{\mu},
p^{a})$, where the components of the spatial dependence $x^{\mu}$
are specified with respect to the global coordinate system. The
momenta $p^{a}$, however, are given in the local vielbein frame.
Then one has, in a kind of mixed representation similar to eq.
(\ref{covboltz}),
\begin{eqnarray}
p^{a}e_{a}{}^{\mu}\frac{\partial f}{\partial
x^{\mu}}+\frac{dp^{b}}{d\lambda}\frac{\partial f}{\partial
p^{b}}=\left[\frac{\delta f}{\delta \lambda}\right]_{coll}\,\,.
\end{eqnarray}

We need to calculate $\frac{dp^{b}}{d\lambda}$. Recalling that the
neutrino trajectories are geodesics fulfilling eq. (\ref{geodesic})
in the global coordinate system, there is

\begin{eqnarray}
\frac{dp^{b}}{d\lambda}
&=& e^{b}{}_{\mu}e^{\mu}{}_{c}\frac{dp^{c}}{d\lambda}\nonumber\\
&=& e^{b}{}_{\mu}\left(\frac{dp^{\mu}}{d\lambda}-\frac{de^{\mu}{}_{c}}{d\lambda}p^{c}\right)\nonumber\\
&=& e^{b}{}_{\mu}\left(-\Gamma^{\mu}{}_{\rho\sigma}p^{\rho}p^{\sigma}-e^{\mu}{}_{c,\rho}p^{\rho}p^{c}\right)\nonumber\\
&=& -e^{b}{}_{\mu}e_{a}{}^{\rho}\left(e^{\mu}{}_{c,\rho}+\Gamma^{\mu}{}_{\rho\sigma}e^{\sigma}{}_{c}\right)p^{a}p^{c}\nonumber\\
&=& -e^{b}{}_{\mu}e_{a}{}^{\rho}e^{\mu}{}_{c;\rho}p^{a}p^{c}\nonumber\\
&=&-\mathfrak{R}^{b}{}_{ac}p^{a}p^{c}\,\,.
\end{eqnarray}
In this formula, the comma denotes a partial derivative and a
semi-colon stands for the covariant derivative. We can define the
so-called Ricci rotation coefficient
\begin{eqnarray}
\mathfrak{R}^{b}{}_{ac}=e^{b}{}_{\mu}e_{a}{}^{\rho}e^{\mu}{}_{c;\rho}\,\,,
\end{eqnarray}
and the Pfaffian derivative (this is the correct partial derivative
in the vielbein frame):
\begin{eqnarray}
\partial_{a}=e_{a}{}^{\mu}\partial_{\mu}\,\,.
\end{eqnarray}
In terms of these quantities the Boltzmann equation eq.
(\ref{covboltz}) becomes
\begin{eqnarray}
p^{a}\left[\partial_{a}-\mathfrak{R}^{b}{}_{ac}p^{c}\frac{\partial}{\partial
p^{b}}\right]f=\left[\frac{\delta f}{\delta
\lambda}\right]_{coll}\,\,. \label{lioucurved}
\end{eqnarray}
This is the general form of the left-hand-side of the Boltzmann
equation (sometimes referred to as the Liouville operator) in curved
spacetime. In non-negligible gravitational fields this Liouville
operator eq. (\ref{lioucurved}) replaces the flat space
left-hand-side of the Boltzmann equation as for example shown in eq.
(\ref{classboltz}).
\section{$4\times4$ Mixing Hamiltonian}
\label{sec:curvedmixing} We increase the dimensionality of the
mixing Hamiltonian employed in flat spacetime, eq.
(\ref{mixhamilton}), to further decompose the phase-space densities
and the associated intensities not only into their flavor degrees of
freedom, but also into their spin degrees of freedom. The matrix
elements of the Wigner phase-space density are then written as a
$4\times4$ matrix:
\begin{eqnarray}
\mathfrak{F}&=&\langle n_{i}|\rho|n_{j}\rangle\nonumber\\
            &=& \left(
      \begin{matrix}
      f_{\nu_{e}}^{\uparrow}      &  f_{e\mu}^{\uparrow}      & f^{\uparrow\downarrow}_{\nu_{e}}& f^{\uparrow\downarrow}_{e\mu} \\
      f_{\mu e}^{\uparrow}        &  f_{\nu_{\mu}}^{\uparrow} & f^{\uparrow\downarrow}_{\mu e}  & f^{\uparrow\downarrow}_{\nu_{\mu}}\\
      f^{\downarrow\uparrow}_{\nu_{e}}& f^{\downarrow\uparrow}_{e\mu}      &  f_{\nu_{e}}^{\downarrow}  &  f_{e\mu}^{\downarrow}\\
      f^{\downarrow\uparrow}_{\mu e}  & f^{\downarrow\uparrow}_{\nu_{\mu}} &  f_{\mu e}^{\downarrow} &  f_{\nu_{\mu}}^{\downarrow}
      \end{matrix}
\right)\,\,. \label{curvedmatrixelements}
\end{eqnarray}
In this matrix $i$ and $j$ run over flavor degrees of freedom and
the spin states. $\uparrow$ denotes left-handed states and
$\downarrow$ means right-handed states. In addition to the
previously employed macroscopic overlap for the differently flavored
neutrinos in the ensemble, eq. (\ref{flavoroverlap}), we must also
consider mixed spin states denoted by the superscripts
$\uparrow\downarrow$ and $\downarrow\uparrow$. Again this is only
relevant if the magnetic moment of the neutrino is comparatively
large and the magnetic fields are very strong (Piriz, Roy \& Wudka
1996). The interaction of the neutrino spinor with the
electromagnetic field stems from terms of the form
\begin{eqnarray}
\mu S^{ab}F_{ab}\psi\,\,,
\end{eqnarray}
where $\psi$ is the neutrino spinor, as for example defined in eq.
(\ref{diraccurved}), and $\mu$ the magnetic moment of the neutrino,
eq. (\ref{magmom}). For $S^{ab}$ we have the expression eq.
(\ref{spinorrep}) from the Appendix \ref{vielbein}:
\begin{eqnarray}
S^{ab}=\frac{i}{4}\left[\gamma^{a},\gamma^{b}\right]\,\,.
\end{eqnarray}
This electromagnetic interaction term can be added to the Dirac
equation (\ref{diraccurvedmod}), yielding
\begin{eqnarray}
\left(i\gamma^{a}e_{a}{}^{\mu}\left[\partial_{\mu}+\left(A_{\mu}+\mathfrak{G}_{\mu}\right)\mathcal{P}_{L}\right]
+\mu S^{ab}F_{ab}+\mathcal{M}\right)\psi=0\,\,.
\label{emdiraccurved}
\end{eqnarray}
In the mixing Hamiltonian, the surviving electromagnetic terms are
$\mu B$ with $B$ being the component of the magnetic field
perpendicular to the neutrino trajectory. Similar to as performed in
the papers by Piriz, Roy \& Wudka 1996, and Br\"uggen 1998 (these
references decomposed the neutrino wavefunction into their spin and
flavor states), we generalize the mixing Hamiltonian of eq.
(\ref{mixhamilton}) as follows:
\begin{eqnarray}
\mathfrak{W}(x^{\mu},p^{\mu})=\mathfrak{W}_{\text{vac}}+\mathfrak{W}_{\text{mat}}(x^{\mu})+
\mathfrak{W}_{\nu\nu}(p^{\mu},x^{\mu})-
\tilde{\mathfrak{W}}_{\nu\nu}(p^{\mu},x^{\mu})+\mathfrak{W}_{\text{grav}}(x^{\mu})\,\,.
\end{eqnarray}
This Hamiltonian includes the weak, electromagnetic and the
gravitational interactions. Adding up all the contributions, except
for the self-interactions, the most general form of the mixing
Hamiltonian is: \vspace{5mm}
\begin{eqnarray}
\mathfrak{W}=\left(
      \begin{matrix}
      \mathcal{V}_{cc}+\mathcal{V}_{nc}+P_{\mu}\mathfrak{G}^{\mu}/\varepsilon-\frac{\Delta m^{2}}{4\varepsilon}\cos2\theta     &  \frac{\Delta m^{2}}{4\varepsilon}\sin2\theta    & \mu_{\nu_{e}}B     & \mu_{e\mu}B \\
      \frac{\Delta m^{2}}{4\varepsilon}\sin2\theta        & \mathcal{V}_{nc}+P_{\mu}\mathfrak{G}^{\mu}/\varepsilon+\frac{\Delta m^{2}}{4\varepsilon}\cos2\theta & \mu_{\mu e}B  & \mu_{\nu_{\mu}}B\\
      \mu_{\nu_{e}}B   & \mu_{e\mu}B       &  -\frac{\Delta m^{2}}{4\varepsilon}\cos2\theta  &   \frac{\Delta m^{2}}{4\varepsilon}\sin2\theta\\
      \mu_{\mu e}B     & \mu_{\nu_{\mu}}B  &   \frac{\Delta m^{2}}{4\varepsilon}\sin2\theta  &  \frac{\Delta m^{2}}{4\varepsilon}\cos2\theta
      \end{matrix}
\right)\,\,.\nonumber\\
\label{mixhamiltoncurved}
\end{eqnarray}

\vspace{5mm}

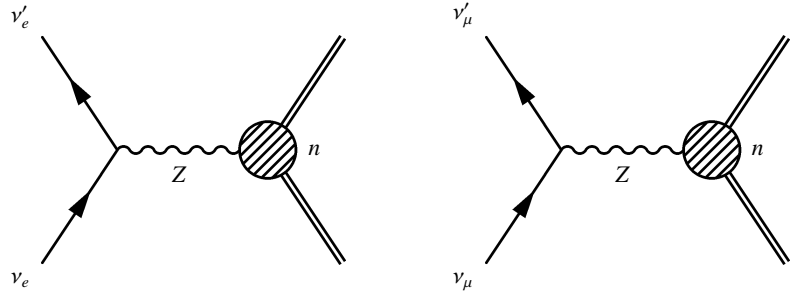
\begin{figure}
\begin{center}
\begin{fmffile}{nc1}
 \begin{fmfgraph*}(50,30)
  \fmfleft{l1,l2}
  \fmfright{r1,r2}
   \fmflabel{$\nu_{e}$}{l1}
   \fmflabel{$\nu_{e}'$}{l2}
   \fmflabel{$\quad n$}{v2}
 \fmf{fermion}{l1,v1,l2}
 \fmf{double}{r1,v2,r2}
  \fmf{boson,label=$Z$}{v1,v2}
  \fmfblob{.15w}{v2}
\end{fmfgraph*}
\end{fmffile}
\hspace{5mm}
\begin{fmffile}{nc2}
 \begin{fmfgraph*}(50,30)
  \fmfleft{l1,l2}
  \fmfright{r1,r2}
   \fmflabel{$\nu_{\mu}$}{l1}
   \fmflabel{$\nu_{\mu}'$}{l2}
   \fmflabel{$\quad n$}{v2}
 \fmf{fermion}{l1,v1,l2}
 \fmf{double}{r1,v2,r2}
  \fmf{boson,label=$Z$}{v1,v2}
  \fmfblob{.15w}{v2}
\end{fmfgraph*}
\end{fmffile}\\[10mm]
\caption{Neutral-current scattering with neutrons for $\nu_{e}$'s
and $\nu_{\mu}$'s.} \label{neutralcurrent}
\end{center}
\end{figure}

$\mathcal{V}_{cc}$ is the charged-current contribution leading to
the effective mass of the $\nu_{e}$ in matter multiply discussed in
this work (the covariant formulation is done in subsection
\ref{subsec:effmass}, and the standard way is treated via the matter
mixing Hamiltonian in eq. (\ref{standardeffmass})):
\begin{eqnarray}
\mathcal{V}_{cc}=2\sqrt{2}G_{f}n_{e}\,\,. \label{electronpotential}
\end{eqnarray}
$\mathcal{V}_{nc}$ is the neutral-current contribution (Mohapatra \&
Pal 2003):
\begin{eqnarray}
\mathcal{V}_{nc}=-2\sqrt{2}G_{f}n_{n}\,\,,
\end{eqnarray}
where $n_{n}$ is the neutron number density and $n_{e}$ is the
electron density. We emphasize that the neutral-current contribution
is the same for all flavors of neutrinos (which is why it was
dropped in eq. (\ref{standardeffmass}), whereas the charged-current
contribution affects the $\nu_{e}$ only (see Fig.
\ref{chargedcurrent})). For neutral-current scattering on the
neutrons in the ambient matter, we have the diagrams depicted in
Fig. \ref{neutralcurrent}.\\

\subsection{Generalized resonances}
\label{gravres}
By virtue of the mixing Hamiltonian eq. (\ref{mixhamiltoncurved}) we
realize that resonances occur whenever the diagonal elements cross.
We read off the resonance conditions for
spin-flips and MSW conversions.\\[3mm]
For $f^{\uparrow}_{\nu_{e}}\leftrightarrow
f^{\uparrow}_{\nu_{\mu}}$:
\begin{eqnarray}
\mathcal{V}_{cc}-\frac{\Delta m^{2}}{2\varepsilon}\cos\theta=0\,\,,
\label{mswtransition}
\end{eqnarray}
for $f^{\uparrow}_{\nu_{e}}\leftrightarrow
f^{\downarrow}_{\nu_{e}}$:
\begin{eqnarray}
\mathcal{V}_{cc}+\mathcal{V}_{nc}+P_{\mu}\mathfrak{G}^{\mu}/\varepsilon=0\,\,,
\end{eqnarray}
for $f^{\uparrow}_{\nu_{e}}\leftrightarrow
f^{\downarrow}_{\nu_{\mu}}$:
\begin{eqnarray}
\mathcal{V}_{cc}+\mathcal{V}_{nc}+P_{\mu}\mathfrak{G}^{\mu}/\varepsilon-\frac{\Delta
m^{2}}{2\varepsilon}\cos2\theta=0\,\,,
\end{eqnarray}
for $f^{\uparrow}_{\nu_{\mu}}\leftrightarrow
f^{\downarrow}_{\nu_{e}}$:
\begin{eqnarray}
\mathcal{V}_{cc}+\mathcal{V}_{nc}+P_{\mu}\mathfrak{G}^{\mu}/\varepsilon+\frac{\Delta
m^{2}}{2\varepsilon}\cos2\theta=0\,\,,
\end{eqnarray}
and lastly for the transition
$f^{\uparrow}_{\nu_{\mu}}\leftrightarrow
f^{\downarrow}_{\nu_{\mu}}$:
\begin{eqnarray}
\mathcal{V}_{nc}+P_{\mu}\mathfrak{G}^{\mu}/\varepsilon=0\,\,.
\end{eqnarray}
In the active-active MSW transition, eq. (\ref{mswtransition}), the
matter potential is due to the charged-current interactions only and
proportional to the electron number density (see eq.
(\ref{electronpotential}). To obtain accurate spectra in general
conditions all these resonances must be taken into account. The
gravitational contribution generates a shift of the resonance
positions. The adiabaticity can also be altered (Br\"uggen 1998).
\section{Kinetic equations}
\label{sec:gravkin}

With the aid of eq. (\ref{lioucurved}), the associated
generalization to curved spacetime including the spinstates of the
the Heisenberg-Boltzmann equation, eq. (\ref{heisboltz}), reads:
\begin{eqnarray}
p^{a}\left[\partial_{a}-\mathfrak{R}^{b}{}_{ac}p^{c}\frac{\partial}{\partial
p^{b}}\right]\mathfrak{F}=-i\left[\mathfrak{W},\mathfrak{F}\right]+\mathcal{C}\,\,.
\label{eq:curvedboltz}
\end{eqnarray}
This equation is the main result of Part II of this thesis. Note
that $\mathfrak{F}$ and $\mathfrak{W}$ are now $4\times4$ matrices.
An analogous equation can be written for antineutrinos. Particles
and antiparticles couple identically to the gravitational field, so
no changes need to be made for the gravitational modifications;
neither for the Liouville operator in curved space, eq.
(\ref{lioucurved}), nor for the gravitational contribution
$\mathfrak{G}$ in eq. (\ref{spineff}) to the mixing Hamiltonian
$\mathfrak{W}$, eq. (\ref{mixhamiltoncurved}).\vspace{3mm}

If we were to componentize them like done above in eq.
(\ref{genericboltz}), we would have to deal with sixteen coupled
equations for the neutrinos without the self-interactions. The issue
of making the equations real-valued can naturally be solved by
defining the real and imaginary parts of the off-diagonal elements
of $\mathfrak{F}$ similarly to eq. (\ref{offdia}). Including the
self-interactions, however, one would have to solve 32 nonlinearly
coupled equations. Clearly, for real applications one would have to
break up the neutrino trajectory to care for the most dominant
effect one at a time.\vspace{3mm}

Eq. (\ref{eq:curvedboltz}) can also be used if one wishes to
generalize eq. (\ref{genericboltz}) to curved space without taking
to account the coupling of the still speculative anomalous magnetic
moment to a very large magnetic field. Then $\mathfrak{F}$ and
$\mathfrak{W}$ reduce to the familiar $2\times2$ matrices given in
eqs. (\ref{matrixelements}) and (\ref{mixhamilton}). In this case
the only effect of the curvature is the standard generalization of
the Liouville operator to curved space as performed in eq.
(\ref{lioucurved}).\vspace{3mm}
\section{Conclusions}
\begin{itemize}

\item The basic fundament of eq. (\ref{eq:curvedboltz}) is the classical, general-relativistic
framework with which we generalized the Liouville operator to curved
spaces in eq. (\ref{lioucurved}).\vspace{3mm}

\item The quantum physics is incorporated by means of the general
mixing Hamiltonian, eq. (\ref{mixhamiltoncurved}). We have included
spin degrees of freedom.\vspace{3mm}

\item The ``master equation'' of Part II, eq. (\ref{eq:curvedboltz})
describes, in principle, the evolution of the quasi-classical
density in phase-space for massive neutrinos with collisions
including all known effects -- from the quantum-mechanical viewpoint
as well as from the gravitational perspective.\vspace{3mm}

\end{itemize}

\chapter[Summary of Part II]{Summary of Part II}
We briefly highlight the main features of our formalism in curved
spacetime and discuss physical applications for which this formalism
might be helpful.
\section{Spin-gravity coupling}
Invariance under general coordinate transformations for spinors
naturally leads to the vielbein fields introduced in Appendix
\ref{vielbein}. In short, these considerations lead to an additional
contribution for the differential operator of the Dirac equation
(\ref{diraccurved}) in terms of the spin connection. However, terms
arising from the spin connection are proportional to the identity
matrix in flavor space and do not lead to additional effects
concerning oscillation between flavor species. Moreover, the spin
connection terms are also diagonal in spin space and act on one spin
orientation of the neutrino only.\\

The most general mixing Hamiltonian, eq. (\ref{mixhamiltoncurved}),
has potentially other off-diagonal contribution stemming from
spin-flips from coupling of the neutrino spinor the an external
magnetic field. It is for this case only that the gravitational
resonances in subsection \ref{gravres} are important. It has been
analyzed elsewhere that, assuming large anomalous magnetic moments,
these effects can be potentially large (Piriz, Roy \& Wudka 1996).
\section{Possible applications}
Combining the oscillation physics in curved spacetime from section
\ref{pheno} with the kinetic generalizations from section
\ref{curvedboltz} yields the master equation of Part II, eqs.
(\ref{eq:curvedboltz}). These equations satisfy all
quantum-mechanical and gravitational considerations for the kinetics
of oscillating neutrinos in curved spacetime.\vspace{3mm}

Potential applications include active galactic nuclei (Piriz, Roy \&
Wudka 1996) and the neutrino fluxes in supernova cores. In practice,
however, one would break up the master equations, eq.
(\ref{eq:curvedboltz}), and solve them isolating the dominant
effect. For supernova neutrinos, for instance, we do not need to
care about collisions at nuclear densities because observable flavor
oscillations are matter-suppressed. There, we must focus on the
collision physics with the ambient matter. In the streaming region
above the neutrino sphere, self-interactions become important and
they must be focussed on. Spin-flips only deserve consideration for
very large magnetic fields and so forth.\vspace{3mm}

To close this summary of Part II, we note that it has been proposed
that the nonlinear evolution in a neutrino background leads to a
speed-up of flavor equilibration processes that could potentially
overshadow the gravitational and all other effects (Saywer 2004 and
2005). We leave this question open for future work.

\part{Conclusions}
\chapter[Conclusions]{Conclusions}
\label{conclusions}
In this thesis, we have derived a generalized set of Boltzmann
equations for real-valued phase-space densities of oscillating
neutrinos interacting with a background medium. The off-diagonal
functions of the Wigner phase-space density representing macroscopic
overlap are explicitly included and serve to couple the flavor
states to reflect neutrino oscillation physics. Conceptually, we
have reduced the time evolution of creation and annihilation
operators to that of real valued phase-space densities without
losing quantum-physical accuracy. Important quantum effects such as
matter-enhanced resonant flavor conversion and quantum decoherence
through matter coupling are correctly incorporated. The generalized
Boltzmann equations are simple and very similar to the equations of
classical transport theory. Neutrino oscillations are incorporated
by new sink and source terms that indirectly couple the expanded set
of equations. Using this formalism, codes that now solve the
standard Boltzmann equations for the classical neutrino phase-space
density ($f_{\nu_i}$), or which address its angular and/or energy
moments, can straightforwardly be reconfigured by the simple
addition of source terms and similar transport equations for overlap
densities that have the same units as $f_{\nu_i}$, to incorporate
neutrino oscillations in a quantum-physically consistent fashion. In
curved spacetime, oscillation phenomenology is altered. These
effects play no observable role except for the usual redshift and
the contribution to the mixing Hamiltonian from the spin connection
can be isolated in the diagonal of the left-handed spin sector of
the effective mixing Hamiltonian. If, however, spin flips occur
through coupling of an anomalous magnetic moment of the neutrino
with a large magnetic field, one must take into account
gravitational resonances similar to the MSW effect in flat
spacetime.

\backmatter%%%%%%%%%%%%%%%%%%%%%%%%%%%%%%%%%%%%%%%%%%%%%%%%%%%%%%%

\appendix
\part{Appendices}
\chapter[Cross section $\mathbf{\nu_{e}+n \longrightarrow
e^{-}+p}$]{Cross section $\mathbf{\nu_{e}+n \longrightarrow
e^{-}+p}$}

\begin{figure}
\begin{center}
\begin{fmffile}{absorb}
 \begin{fmfgraph*}(50,30)
  \fmfleft{l1,l2}
  \fmfright{r1,r2}
   \fmflabel{$\nu_{e}$}{l1}
   \fmflabel{$e^{-}$}{l2}
   \fmflabel{$n$}{r1}
   \fmflabel{$p$}{r2}
 \fmf{fermion}{l1,v1,l2}
 \fmf{double}{r1,v2,r2}
  \fmf{boson,label=$W$}{v1,v2}
  \fmfblob{.15w}{v2}
\end{fmfgraph*}
\end{fmffile}\\[6mm]
\caption{Super--allowed charged--current $\nu_{e}$ absorption on
neutrons.} \label{absorb}
\end{center}
\end{figure}
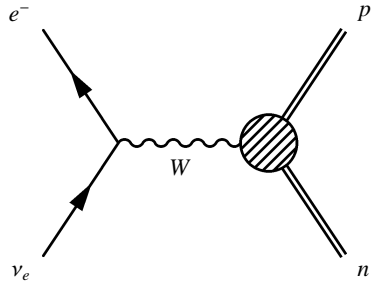

%\addcontentsline{toc}{chapter}{Appendix}
%
\label{appendixcross}
A convenient reference neutrino cross section is $\sigma_o$, given
by
\begin{eqnarray}
\sigma_o\,=\,\frac{4G_F^2(m_ec^2)^2}{\pi(\hbar c)^4}\simeq
\,1.705\times 10^{-44}\,\text{cm}^2\,\,,
\end{eqnarray}
where $G_F$ is the Fermi weak coupling constant ($\simeq 1.436\times
10^{-49}$ ergs cm$^{-3}$). The total $\nu_{e}-n$ absorption cross
section for the reaction $\nu_{e}+n \longrightarrow e^{-}+p$ is then
given by
\begin{eqnarray}
\sigma^a_{\nu_{e}n}\sim\,\sigma_{0}\left(\frac{1+3g_A^2}{4}\right)\,\left(\frac{\varepsilon_{\nu_e}+\Delta_{np}}{m_ec^2}\right)^2\,
\left[1-\left(\frac{m_ec^2}{\varepsilon_{\nu_e}+\Delta_{np}}\right)^2\right]^{1/2}\,\,
, \label{ncapture}
\end{eqnarray}
where $g_A$ is the axial--vector coupling constant ($\sim -1.23$),
and $\Delta_{np}=m_nc^2-m_pc^2=1.29332$ MeV . In Fig. (\ref{absorb})
we show the heuristic Feynman graph corresponding to this
process.\\
\chapter[Vielbein (Tetrad) Formalism and Spinor Fields]{Vielbein (Tetrad) Formalism and Spinor Fields}
\label{vielbein}
The essential quantities in the general theory of relativity such as
the curvature or the energy-momentum are invariant under general
coordinate transformations. This can be viewed as the construction
principle of the general theory of relativity (Weinberg 1972).
Luckily, these quantities behave like tensors not only under general
coordinate transformation but they also behave as tensors under
Lorentz transformations. This makes the transition from special
relativity in flat spacetime to general relativity in curved
spacetimes straightforward. For spinor fields this is not true
because the tensor representations of the group GL(4) of general
linear $4\times4$ matrices behave like tensors under the subgroup of
Lorentz transformations, but there are no representation of GL(4)
which behave like spinors under the Lorentz subgroup. To remedy this
situation, one erects at every point $x$ a noncoordinate basis
$\xi^{a}(x)$ that is locally inertial. Then the metric can be
written
\begin{eqnarray}
g_{\mu\nu}(x)=e_{\mu}{}^{a}(x)e_{\nu}{}^{b}(x)\eta_{ab}\,\,,
\end{eqnarray}
where for a Lorentzian spacetime $\eta_{ab}$ represents the
Minkowski metric, while in a space with positive-definite metric it
would represent the Euclidean metric. The Latin indices indicate
that the locally inertial coordinates are not related to any
coordinate system. The vielbeins are given by
\begin{eqnarray}
e{}^{a}_{\mu}(x)=\left(\frac{\partial \xi^{a}(x')}{\partial
x^{\mu}}\right)_{x'=x}\,\,. \label{viel}
\end{eqnarray}
Coordinate tensors of arbitrary rank can be made into a set of
scalars by expressing their components with respect to the locally
inertial basis, that is,
\begin{eqnarray}
T^{a}{}_{b}=e^{a}{}_{\mu}e_{b}{}^{\nu}T^{\mu}{}_{\nu}\,\,.
\end{eqnarray}
The point is that now, just as in global Minkowski space, the change
of the locally inertial basis in the vielbein frame corresponds to a
(now spatially dependent) Lorentz transformation. The two
transformations, local Lorentz transformations and general
coordinate transformations, can be executed at the same time
resulting in the mixed transformation law (Carroll 2004):
\begin{eqnarray}
T^{a'\mu'}{}{}_{b'\nu'}=\Lambda^{a'}{}_{a}\frac{\partial
x^{\mu'}}{\partial x^{\mu}}\Lambda^{b}{}_{b'}\frac{\partial
x^{\nu}}{\partial x^{\nu'}}T^{a\mu}{}{}_{b\nu}\,\,.
\end{eqnarray}

In general an arbitrary field in the vielbein frame $\psi_{n}(x)$
transforms as
\begin{eqnarray}
\psi_{n}(x)=\sum_{m}\left[D\left(\Lambda(x)\right)\right]_{nm}\psi_{m}(x)\,\,,
\end{eqnarray}
where $D(\Lambda)$ is a matrix representation of the Lorentz group
tailored to the nature of the field. Heuristically, we have stored
away the curvature in the vielbeins so as to now being able, for a
spinor field, to apply the spinor representation of the Lorentz
group.\\

We now need a derivative operator, $\mathfrak{D}_{a}$, that is a
scalar with respect to the general coordinate transformations and a
Lorentz vector with respect to local position-dependent Lorentz
transformations with matrix representation $\Lambda_{a}{}^{b}(x)$.
This ensures, that any action which depends on the various fields of
interest, including spinor fields, will then automatically be
independent of the choice of locally inertial frames if it is
invariant under ordinary constant Lorentz transformations. The
derivative operator must transform homogeneously:
\begin{eqnarray}
\mathfrak{D}_{a}\psi(x)\rightarrow\Lambda_{a}{}^{b}(x)D\left(\Lambda(x)\right)
\mathfrak{D}_{b}\psi(x)\,\,. \label{derivop}
\end{eqnarray}
It can be shown that the expression following expression is the
correct derivative operator:
\begin{eqnarray}
\mathfrak{D}_{a}=e_{a}{}^{\mu}\left(\partial_{\mu}+\Gamma_{\mu}\right)\,\,,
\end{eqnarray}
where the spin connection $\Gamma_{\mu}$ is given by (Weinberg
1972):
\begin{eqnarray}
\Gamma_{\mu}=\frac{1}{2}S^{bc}e_{b}{}^{\nu}e_{c\nu;\mu}\,\,,
\end{eqnarray}
where the semicolon denotes the covariant derivative, and $S^{bc}$
is the appropriate representation of the Lorentz group. For spinor
fields, employing the chiral choice of $\gamma$'s, eq.
(\ref{chiral}), one has
\begin{eqnarray}
S^{bc}=\frac{i}{4}\left[\gamma^{b},\gamma^{c}\right]\,\,,
\label{spinorrep}
\end{eqnarray}
where square brackets denote the usual matrix commutators.\\[3mm]

To summarize: The effects of gravitation on any physical system can
be taken into account by writing down the matter action or the field
equations that hold in special relativity, and then replacing all
derivatives $\partial/\partial x^{a}$ with the derivative operator
$\mathfrak{D}_{a}$ of eq. (\ref{derivop}). For the case of
consistently describing spinor fields in curved space one must
resort to this line of thought, since only in the vielbein frame can
one employ the spinor representations of the Lorentz group
coherently.

 \noindent
 \small Abazajian, K., Bell, N., Fuller, G. M., and Wong, Y. Y. Y.: astro-ph/0410175 (2004).\\[4mm]
 Abazajian, K., Fuller, G. M., and Patel, M.: Phys. Rev. D \textbf{64}, 023501 (2001).\\[4mm]
 Akhiezer, A. I., and Peletminskii, S. V.: \textit{Methods of Statistical Physics} (Pergamon Press, Volume 104, 1981).\\[4mm]
 Akhmedov, E., Barroso, A., and Keraenen, P.: Phys. Lett. B \textbf{517}, 355 (2001).\\[4mm]
 Bell, N. F., Rawlinson, A. A., and Sawyer, R. F.: Phys. Lett. B \textbf{573} 86-93 (2003).\\[4mm]
 Bethe, H.: Phys. Rev. Lett. \textbf{56} 1305 (1986).\\[4mm]
 Bilenky, S. M., and Petcov, S. T.: Rev. Mod. Phys. \textbf{59}, 671 (1987).\\[4mm]
 Bilenky, S. M., and B. Pontecorvo, Phys. Lett. B \textbf{61}, 248 (1976).\\[4mm]
 Boyanovsky, D., and Ho, C. M.: Phys. Rev. D \textbf{69}, 125012 (2004).\\[4mm]
 Brill, D. R, and Wheeler, J. A.: Rev. Mod. Phys. \textbf{29}, 3 (1957).\\[4mm]
 Br\"uggen, M.: Phys. Rev. D \textbf{58}, 083002 (1998).\\[4mm]
 Burrows, A., Gandhi, R.: Phys. Lett. B \textbf{243}, 149 (1995).\\[4mm]
 Burrows, A., Young, T., Pinto, P., Eastman, R., and Thompson T. A.: Astrophys. J. \textbf{539}, 865 (2000).\\[4mm]
 Burrows, A., Reddy, S., and Thompson, T. A.: astro-ph/0404432 (2004).\\[4mm]
 Cardall, C. Y., and Fuller, G. M.: Phys. Rev. D \textbf{55}, 7960 (1997).\\[4mm]
 Carroll, S. M.: \textit{Spacetime and Geometry}, 1st edn (Addison Wesley 2004).\\[4mm]
 de Groot, S., van Leeuwen, W., and van Weert, C. G.: \textit{Relativistic Kinetic Theory} (North Holland, 1980).\\[4mm]
 Dolgov, A., Barbieri, R.: Nucl. Phys. B \textbf{349}, 743-753 (1990).\\[4mm]
 Friedland, A., and Lunardini, C.: JHEP10 043 (2003).\\[4mm]
 Friedland, A., and Lunardini, C.: Phys. Rev. D \textbf{68}, 013007 (2003).\\[4mm]
 Fukuda, S., \etal : Phys. Lett. B \textbf{539}, 179 (2002).\\[4mm]
 Giunti, C.: \textit{Neutrino Mixing and Oscillations}, www.nu.to.infn.it/ (2004).\\[4mm]
 Lambiase, G.: astro-ph/0412408 (2004).\\[4mm]
 Lambiase, G., Papini, G., Punzi, R. and Scarpetta, G.: gr-qc/0503027 (2005).\\[4mm]
 Kayser, B.: Phys. Rev. D \textbf{24}, 110 (1981).\\[4mm]
 Landau, L. D.: Sov. Phys. JETP \textbf{5}, 101 (1957).\\[4mm]
 McKellar, B. H. J., Thomson, M. J.: Phys. Rev. D \textbf{51}, 4 (1992).\\[4mm]
 Mihalas, D., and Weibel-Mihalas, B.: \textit{Foundations of Radiation Hydrodynamics} (Dover Publications, 1999).\\[4mm]
 Mikheev, S., and Smirnov, A. Y.: Sov. Phys. JETP \textbf{64}, 4 (1986).\\[4mm]
 Moertsell, E., Bergstroem, L., and Goobar, A.: Phys. Rev. D \textbf{37}, 1237 (2002).\\[4mm]
 Mohapatra R. N, and Pal, P. B.: \textit{Massive Neutrinos in Phys. and Astrophys.} (World Scientific, 2003).\\[4mm]
 Pantaleone, J., and Kuo T. K.: Rev. of Mod. Phys. \textbf{61}, 937 (1989).\\[4mm]
 Pantaleone, J.: Phys. Lett. B \textbf{287}, 128 (1992) A.\\[4mm]
 Pantaleone, J.: Phys. Rev. D \textbf{46}, 510 (1992) B.\\[4mm]
 Pantaleone, J.: Phys. Lett. B \textbf{342}, 250-256 (1995).\\[4mm]
 Parke, S. J.: Phys. Rev. Lett. \textbf{57}, 1275 (1986).\\[4mm]
 Peskin, M. E., Schroeder D. V.: \textit{An Introduction to Quantum Field Theory} (Westview Press, 1995).\\[4mm]
 Piriz, D., Roy, M. and Wudka J.: Phys. Rev. D \textbf{54}, 1587 (1996).\\[4mm]
 Qian, Y.-Z., and Fuller, G.: Phys. Rev. D. \textbf{20} (1995).\\[4mm]
 Raffelt, G.: \textit{Stars as Laboratories for Fundamental Physics} (The University of Chicago Press, 1996).\\[4mm]
 Raffelt, G., and Sigl, G.: Nucl. Phys. B \textbf{406}, 423 (1993).\\[4mm]
 Raffelt, G., and Sigl, G.: Astropart. Phys. \textbf{1}, 165 (1993).\\[4mm]
 Raffelt, G., and Stodolsky, L.: Phys. Rev. D. \textbf{37}, 1237 (1988).\\[4mm]
 Raffelt, G., Sigl, G., and Stodolsky, L.: Phys. Rev. D \textbf{45}, 1782 (1992).\\[4mm]
 Raffelt, G., Sigl, G., and Stodolsky, L.: Phys. Rev. Lett. \textbf{70}, 16 (1993).\\[4mm]
 Robinson, D. C.: Phys. Rev. Lett. \textbf{34}, 901 (1975).\\[4mm]
 Rudszky, M. A.: Astrophys. Space Sci. \textbf{165}, 65 (1990).\\[4mm]
 Sawyer, R. F.: Phys. Rev. D \textbf{42}, 3908 (1990).\\[4mm]
 Sawyer, R. F.: hep-ph/0408265 (2004).\\[4mm]
 Sawyer, R. F.: hep-ph/0503013 (2005).\\[4mm]
 Shrock, R. E., and Fujikawa, K.: Phys. Rev. D \textbf{16}, 1444 (1977).\\[4mm]
 Stewart, J. M.: \textit{Non-Equilibrium Relativistic Kinetic Theory} (Springer-Verlag, 1971).\\[4mm]
 Strack, P., and Burrows, A.: Phys. Rev. D \textbf{71}, 093004 (2005).\\[4mm]
 Takahashi, K., Sato, K., Burrows, A., and Thompson, T. A.: Phys. Rev. D. \textbf{68}, 113009 (2003).\\[4mm]
 Thompson, T. A., Burrows, A., and Horvath, J. E.: Phys. Rev. C. \textbf{62}, 035802 (2000).\\[4mm]
 Weinberg, S. W.: \textit{Gravitation and Cosmology} (Wiley, New York, 1972).\\[4mm]
 Wigner, E.: Phys. Rev. \textbf{40}, 749 (1932).\\[4mm]
 Wigner, E., Hillary, M., O'Connell, R. F., and Scully, M. O.: Phys. Rep. \textbf{106}, 121 (1984).\\[4mm]
 Wolfenstein, L.: Phys. Rev. D \textbf{17} (1978).\\[4mm]
 Wolfenstein, L.: Phys. Rev. D \textbf{20}, 2634 (1979).\\[4mm]

%

%\printindex

%%%%%%%%%%%%%%%%%%%%%%%%%%%%%%%%%%%%%%%%%%%%%%%%%%%%%%%%%%%%%%%%%%%%%%

\end{document}